\documentclass[conference]{IEEEtran}

\usepackage{adjustbox}

\usepackage{cite}
\usepackage{hyperref}
\ifCLASSINFOpdf
\else
\fi
\usepackage[cmex10]{amsmath}
\usepackage{amsmath,amsfonts,amsthm}
\usepackage{amssymb}
\usepackage{cleveref}
\theoremstyle{plain}
\newtheorem{theorem}{Theorem}
\newtheorem{assumption}{Assumption}
\newtheorem{corollary}{Corollary}
\newtheorem{lemma}{Lemma}
\newtheorem{remark}{Remark}

\usepackage{mathtools}

\DeclareMathAlphabet{\mathcal}{OMS}{cmsy}{m}{n}
\usepackage{algpseudocode}
\usepackage[ruled, lined, linesnumbered, commentsnumbered, longend]{algorithm2e}
\usepackage{ifthen}
\usepackage{xcolor}
\definecolor{myblue}{rgb}{0,0,0.6}

\SetCommentSty{mycommfont}
\usepackage{array}
\usepackage{mdwmath}
\usepackage{mdwtab}
\usepackage{eqparbox}
\usepackage{balance}

\usepackage{caption}
\usepackage{subcaption}
\usepackage{booktabs}
\usepackage{multirow}

\usepackage{stfloats}
\usepackage{url}
\usepackage{xargs}
\usepackage{xcolor}
\definecolor{MidnightBlue}{rgb}{0.1, 0.1, 0.44}
\definecolor{CornflowerBlue}{rgb}{0.39, 0.58, 0.93}
\definecolor{shadecolor}{rgb}{0.8,0.8,0.8}
\definecolor{Black}{rgb}{0.0, 0.0, 0.0}
\definecolor{BrickRed}{rgb}{0.8, 0.25, 0.33}

\ifx\commentsoff\undefined

\fi
\usepackage[normalem]{ulem}
\newsavebox{\imagebox}
\usepackage{ccicons}

\begin{document}
\title{Securing Federated Sensitive Topic Classification against Poisoning Attacks}

%\author{\IEEEauthorblockN{Michael Shell}
%\IEEEauthorblockA{Georgia Institute of Technology\\
%someemail@somedomain.com}
%\and
%\IEEEauthorblockN{Homer Simpson}
%\IEEEauthorblockA{Twentieth Century Fox\\
%homer@thesimpsons.com}
%\and
%\IEEEauthorblockN{James Kirk\\ and Montgomery Scott}
%\IEEEauthorblockA{Starfleet Academy\\
%someemail@somedomain.com}}

\author{
\IEEEauthorblockN{Tianyue Chu}
\IEEEauthorblockA{IMDEA Networks Institute\\
Universidad Carlos III de Madrid}
%tianyue.chu@imdea.org}
\and

\IEEEauthorblockN{Alvaro Garcia-Recuero}
\IEEEauthorblockA{IMDEA Networks Institute}
%someemail@somedomain.com}
\and
\IEEEauthorblockN{Costas Iordanou}
\IEEEauthorblockA{Cyprus University of Technology}
%someemail@somedomain.com}

\newlineauthors
\IEEEauthorblockN{Georgios Smaragdakis}
\IEEEauthorblockA{TU Delft}
%someemail@somedomain.com}
\and
\IEEEauthorblockN{Nikolaos Laoutaris}
\IEEEauthorblockA{IMDEA Networks Institute}
%nikolaos.laoutaris@imdea.org}
}


%Telephone: (800) 555--1212, Fax: (888) 555--1212}


\IEEEoverridecommandlockouts
\makeatletter\def\@IEEEpubidpullup{6.5\baselineskip}\makeatother
\IEEEpubid{\parbox{\columnwidth}{
    Network and Distributed System Security (NDSS) Symposium 2023\\
   27 February - 3 March 2023, San Diego, CA, USA\\
   ISBN 1-891562-83-5\\
   https://dx.doi.org/10.14722/ndss.2023.23112\\
   www.ndss-symposium.org\\
    Attribution-NonCommercial-NoDerivatives 4.0 International\\
    (CC BY-NC-ND 4.0)\ccbyncnd
}
\hspace{\columnsep}\makebox[\columnwidth]{}}

%Command for two line authors
\makeatletter
\newcommand{\newlineauthors}{%
  \end{@IEEEauthorhalign}\hfill\mbox{}\par
  \mbox{}\hfill\begin{@IEEEauthorhalign}
}
\makeatother

\maketitle

\begin{abstract}
%\boldmath
We present a Federated Learning (FL) based solution for building a distributed classifier capable of detecting URLs containing sensitive content, i.e., content related to categories such as health, political beliefs, sexual orientation, etc. Although such a classifier addresses the limitations of previous offline/centralised classifiers, it is still vulnerable to poisoning attacks from malicious users that may attempt to reduce the accuracy for benign users by disseminating faulty model updates. To guard against this, we develop a robust aggregation scheme based on subjective logic and residual-based attack detection. Employing a combination of theoretical analysis, trace-driven simulation, as well as experimental validation with a prototype and real users, we show that our classifier can detect sensitive content with high accuracy, learn new labels fast, and remain robust in view of poisoning attacks from malicious users, as well as imperfect input from non-malicious ones.
\end{abstract}


% \fi
%%\IEEEpeerreviewmaketitle

\section{Introduction}

Most people are not aware  that tracking services are present even on sensitive web domains.
%The web contains many domains in which most people %would rather not be seen by third-party tracking %services. 
%Indeed, 
Being tracked on a cancer discussion forum, a dating site, or a news site with non-mainstream political affinity can be considered an ``elephant in the room'' when it comes to the anxieties that many people have about their online privacy. %\ci{we can move the following definition in the abstract since we already define FL there as well.} 
The General Data Protection Regulation (GDPR)~\cite{EU-GDPR} puts specific restrictions on the collection and processing of sensitive personal data \textit{``revealing racial or ethnic origin, political opinions, religious or philosophical beliefs, or trade union membership, also genetic data, biometric data for the purpose of uniquely identifying a natural person, data concerning health or data concerning a natural persons sex life or sexual orientation''}. So do other public bodies around the world, \textit{e.g.} in California (California Consumer Privacy Act (CCPA)~\cite{CCPA}), Canada~\cite{Canada-privacy}, Israel~\cite{Israel-privacy}, Japan~\cite{JPIP}, and Australia~\cite{Australia-privacy}.

%\vspace{-0.5em} 
In a recent paper, Matic et al.~\cite{matic2020identifying} showed how to train a classifier for detecting whether the content of a URL relates to any of the above-mentioned sensitive categories. The classifier was trained using 156 thousand sensitive URLs obtained from the Curlie~\cite{Curlie} crowdsourced web taxonomy project. Despite the demonstrated high accuracy, this method has limitations that stem from being \emph{centralised} and \emph{tied to a fixed training set}. 
The first limitation means that the method cannot be used ``as is'' to drive a privacy-preserving distributed classification system. The second limitation implies that it is not straight-forward to cover new labels related to yet unseen sensitive content. For example, in their work the Health category could be classified with accuracy greater than 90\%. However, the training labels obtained from Curlie in 2020 did not include any labels related to the COVID-19 pandemic. Therefore, as will be shown later, this classifier classifies COVID-19 related sites 
%that contain sensitive health content 
with only 53.13\% accuracy. 

%\vspace{-.5em}
Federated Learning (FL)~\cite{mcmahan2017communication, konevcny2016federated} %\ci{we already define FL in Abstract.} 
offers a natural solution to the above two mentioned limitations, namely, centralized training and training for a fixed training set. %since 
FL allows different clients to train their classification models locally without revealing new or existing sensitive URLs that they label, while collaborating by sharing model updates that can be combined to build a superior global classification model. FL has proved its value in a slew of real-world applications, ranging from mobile computing~\cite{hard2018federated, feng2020pmf, yu2020learning} to health and medical applications~\cite{huang2019patient, brisimi2018federated, lee2018privacy}. However, due to its very nature, FL is vulnerable to so-called \emph{poisoning attacks}~\cite{bhagoji2019analyzing, bagdasaryan2020backdoor} mounted by malicious clients that may intentionally train their local models with faulty labels or backdoor patterns, and then disseminate the resulting updates with the intention of reducing the classification accuracy for other benign clients. State-of-the-art approaches for defending against such attacks depend on robust aggregation~\cite{blanchard2017machine, yin2018byzantine, fu2019attack, fung2018mitigating, fung2020limitations, cao2021fltrust} which, as we will demonstrate later, are slow to converge, thereby making them impractical for the sensitive-content classification problem that we tackle in this paper. 
%\vspace{2pt}

%\vspace{-.5em}
\noindent \textbf{Our Contributions:} In this paper, we employ FL 
%for the first time 
for sensitive content classification. We show how to develop a robust FL method for classifying arbitrary URLs that may contain GDPR sensitive content. Such a FL-based solution allows building a distributed classifier that can be offered to end-users in the form of a web browser extension in order to: 
(i) warn them before and while they navigate into such websites, especially when they are populated with trackers, and (ii) allow them to contribute new labels, e.g., health-related websites about COVID-19, and thus keeping the classifier always up-to-date. To the best of our knowledge this method represents the first use of FL for such task.
%Federated Learning \ci{FL?} 

%\vspace{-.5em}
Our second major contribution is the development of a reputation score for protecting our FL-based solution from poisoning attacks~\cite{bagdasaryan2020backdoor, bhagoji2019analyzing}. Our approach is based on a novel combination of subjective logic~\cite{josang2016subjective} with residual-based attack detection. %\vspace{-0.5em}
Our third contribution is the development of an extensive theoretical and experimental performance evaluation framework for studying the accuracy, convergence, and resilience to attacks of our proposed mechanism. %\vspace{-0.5em}
Our final contribution is the implementation of our methods in a prototype system called \emph{EITR} (standing for ``Elephant In the Room'' of privacy) and our preliminary experimental validation with real users tasked to provide  fresh labels for the accurate classification of COVID-19 related URLs. 
%\vspace{2pt}

\noindent \textbf{Our findings:} Using a combination of theoretical analysis, simulation, and experimentation with real users, we:
%\vspace{-2mm}
\begin{itemize}
[
    \setlength{\IEEElabelindent}{\dimexpr-\labelwidth-\labelsep}% Wrapping of text beyond first line of \item
    \setlength{\itemindent}{\dimexpr\labelwidth+\labelsep}% identation for each new \item
    \setlength{\listparindent}{\parindent}% Restore regular paragraph indentation
]
\setlength{\itemsep}{0pt}
    \item Demonstrate experimentally that our FL-based classifier achieves comparable accuracy with the centralised one presented in~\cite{matic2020identifying}.
    \item Prove analytically that under data poisoning attacks, our reputation-based robust aggregation built around subjective logic, converges to a near-optimal solution of the corresponding Byzantine fault tolerance problem under standard assumptions. The resulting performance gap is determined by the percentage of malicious users.
    \item Evaluate experimentally our solution against state-of-the-art algorithms such as Federated Averaging~\cite{mcmahan2017communication}, Coordinate-wise median~\cite{yin2018byzantine}, Trimmed-mean~\cite{yin2018byzantine}, FoolsGold~\cite{fung2018mitigating,fung2020limitations}, Residual-based re-weighting~\cite{fu2019attack} and FLTrust~\cite{cao2021fltrust}, and show that our algorithm is robust under Byzantine attacks by using different real-world datasets. We demonstrate that our solution outperforms these popular solutions in terms of convergence speed by a factor ranging from $1.6\times$ to $2.4\times$ while achieving the same or better accuracy. Furthermore, our method yields the most consistent and lowest Attack Success Rate (ASR), with at least 72.3\% average improvement against all other methods.
    \item Validate using our \emph{EITR} browser extension that our FL-based solution can quickly learn to classify health-related sites about COVID-19, even in view of noisy/inconsistent input provided by real users.  
\end{itemize}
%\vspace{-2mm}

The remainder of the article is structured as follows: Section~\ref{sec:Background} introduces the background for our topic. Section~\ref{sec:FL} presents our reputations scheme for FL-based sensitive content classification, as well as its theoretical analysis and guarantees. Section~\ref{sec:performance} covers our extensive performance evaluation against the state-of-the-art and Section~\ref{sec:EITR} some preliminary results from our \emph{EITR} browser extension. Section~\ref{sec:conclusions} concludes the paper and points to on-going and future work including the generalization of our method to other topics.

\section{Background}\label{sec:Background}
\subsection{A Centralised Offline Classifier for Sensitive Content}
Matic et al.~\cite{matic2020identifying} have shown how to develop a text classifier able to detect URLs that contain sensitive content. This classifier is centralised and was developed in order to conduct a one-off offline study aimed at estimating the percentage of the web that includes such content. Despite achieving an accuracy of at least 88\%, utilising a high-quality training set meticulously collected by filtering the Curlie web-taxonomy project~\cite{Curlie}, this classifier cannot be used ``as is'' to protect real users visiting sensitive URLs populated by tracking services.

\subsection{Challenges in Developing a Practical Classifier for Users}
%We apply Federated Learning , for the first time, to the problem of sensitive website classification, which has some unique characteristics and challenges.
%\ci{The following paragraph can be compacted to save space.}
\noindent\textbf{From offline to online:} The classifier in~\cite{matic2020identifying} was trained using a dataset of 156 thousand sensitive URLs. Despite being the largest dataset of its type in recent literature, this dataset is static and thus represents sensitive topics up to the time of its collection. This does not mean, of course, that a new classifier trained with this data would never be able to accurately classify new URLs pertaining to those sensitive categories. This owes to the fact that categories such as Health, involve content and terms that do not change radically with time. Of course, new types of sensitive content may appear that, for whatever reason, may not be so accurately classified using features extracted from past content of the same sensitive category. Content pertaining to the recent COVID-19 pandemic is such an example. Although the Health category had 74,764 URLs in the training set of~\cite{matic2020identifying} which lead to a classification accuracy of 88\% for Health, as we will see later in Figure~\ref{fig:real_user}~middle of Section~\ref{sec:covid-validation}, the classifier of~\cite{matic2020identifying} classifies accurately as Health only 53.13\% of the COVID-19 URLs with which we tested it. This should not come as a surprise since the dataset of~\cite{matic2020identifying} corresponds to content generated before the first months of 2020, during which COVID-19 was not yet a popular topic. Therefore, we need to find a way to update an existing classifier so that it remains accurate as new sensitive content appears.   

%\vspace{-0.5em}
\noindent\textbf{From centralised to distributed:} A natural way to keep a classifier up-to-date is to ask end-users to label new sensitive URLs as they encounter them. End-users can report back to a centralised server such URLs which can then be used to retrain the classification model. This, however, entails obvious privacy challenges of ``Catch-22'' nature, since to protect users by warning them about the presence of trackers on sensitive URLs, they would first be required to report to a potentially untrusted centralised server that they visit such URLs. Even by employing some methods for data scarcity, e.g., semi-supervised learning, the manual labelling from users remains sensitive and may be harmed by the untrusted server. Federated Learning, as already mentioned, is a promising solution for avoiding the above Catch22 by conducting a distributed, albeit, privacy-preserving, model training. In an FL approach to our problem, users would label new URLs locally, e.g., a COVID-19 URL as Health, retrain the classifier model locally, and then send model updates, not labelled data, to a centralised server that collects such updates from all users, compiles and redistributes the new version of the model back to them. In Section~\ref{sec:FL} we show how to develop a distributed version of the sensitive topic classifier of~\cite{matic2020identifying} using FL. The trade-off of using FL, is that the distributed learning group becomes vulnerable to attacks, such as ``label-flipping'' poisoning attacks discussed in Section~\ref{sec:performance}. This paper develops a reputation scheme for mitigating such attacks. Other types of attacks and measures for preserving the privacy of users that participate in a FL-based classification system for sensitive content are discussed in Section~\ref{sec:conclusions}.

\subsection{Related Work}

\noindent\textbf{Privacy preserving crowdsourcing:} Similar challenges to the ones discussed in the previous paragraph have been faced in services like the \emph{Price \$heriff}~\cite{iordanou2017fiddling} and \emph{eyeWnder}~\cite{iordanou2019beyond} that have used crowdsourcing to detect online price discrimination and targeted advertising, respectively. Secure Multi-Party Computation (SMPC) techniques such as private $k$-means~\cite{su2016differentially} 
%and count-min sketches~\cite{cormode2005improved} 
are used to allow end-users to send data in a centralised server in a privacy-preserving manner. The centralised computation performed by \emph{Price \$heriff} and \emph{eyeWnder} is not of ML nature, thus leaving data anonymisation as the main challenge, for which SMPC is a good fit. Classifying content as sensitive or not is a more complex ML-based algorithm for which FL is a more natural solution than SMPC.
\\
\noindent\textbf{General works on FL:} 
FL~\cite{mcmahan2017communication, konevcny2016federated} %
is a compelling technique for training large-scale distributed machine learning models while maintaining security and privacy. The motivation for FL is that local training data is always kept by the clients and the server has no access to the data. Due to this benefit that alleviates privacy concerns, several corporations have utilised FL in real world services. In mobile devices, FL is used to predict keyboard input~\cite{hard2018federated}, human mobility~\cite{feng2020pmf} and behaviour for the Internet of Things~\cite{yu2020learning}. FL is also applied in healthcare to predict diseases~\cite{huang2019patient, brisimi2018federated}, detect patient similarity~\cite{lee2018privacy} while overcoming any privacy constrains. For the classification, FL is not only implemented for image classification~\cite{ahn2019unsupervised} but also text classification~\cite{lin2021fednlp}.
\\
\noindent\textbf{Resilience to poisoning attacks:} Owing to its nature~\cite{bhagoji2019analyzing, bagdasaryan2020backdoor}, FL is vulnerable to poisoning attacks, such as label flipping~\cite{fu2019attack} and backdoor attacks~\cite{bagdasaryan2020backdoor}. Therefore, several defence methods have been developed~\cite{yin2018byzantine, fu2019attack, fung2018mitigating,fung2020limitations}. While these state-of-the-art approaches perform excellently in some scenarios, they are not without limitations. First, they are unsuitable for our sensitive content classification, which necessitates that a classifier responds very fast to ``fresh'' sensitive information appearing on the Internet. In existing methods, the primary objective is to achieve a high classification accuracy. This is achieved via statistical analysis of client-supplied model updates and discarding of questionable outliers before the aggregation stage. However, since the server distrusts everyone by default, even if an honest client discovers some fresh sensitive labels, its corresponding updates may be discarded or assigned low weights, up until more clients start discovering these labels. This leads to a slower learning rate for new labels. 
%Thus, these techniques are insufficient for the GDPR-sensitive classification task which necessitates the classifier to categorise newly published contents.
% The reason for this is because the server doesn't trust anyone in the system, so it would measure the quality of the updates in every round to decide to keep it or leave it without considering the historical performance of the clients.

Second, recent studies~\cite{bhagoji2019analyzing, bagdasaryan2020backdoor} have shown that existing Byzantine-robust FL methods are still vulnerable to local model poisoning since they are forgetful by not tracking information from previous aggregation rounds. Thus, an attacker can efficiently mount an attack by spreading it across time~\cite{guerraoui2018hidden}. For example, \cite{karimireddy2021learning} recently showed that even after infinite training epochs, any aggregation which is neglectful of the past cannot converge to an efficient solution. 

%\vspace{-1m}
The preceding studies demonstrate the importance of incorporating clients' previous long-term performance in evaluating their trustworthiness. Few recent studies have considered this approach~\cite{cao2021fltrust,karimireddy2021learning}. In~\cite{karimireddy2021learning}, the authors propose leveraging historical information for optimisation, but not for assessing trustworthiness. In~\cite{cao2021fltrust}, 
%the authors assign 
a trust score is assigned to each client model update according to the cosine similarity between the client's and server's model updates, which is trained on the server's root dataset (details in Section~\ref{subsec:evaluatedmethods}).
However, it is impractical for a server to obtain additional data, such as a root dataset, in order to train a server-side model. In addition, because the server collects root data only once and does not update it throughout the training process, when new types of content emerge over time, the root data may become stale thereby harming the classifier's performance. Other recent studies employ spectral analysis~\cite{shejwalkar2021manipulating}, differential privacy~\cite{naseri2020local}, and deep model inspection~\cite{rieger2022deepsight} to guard against poisoning attacks, but, again, they do not use historical information to assess the reliability of clients.
To measure client trustworthiness without collecting additional data at the server, in the next sections we show how to design a robust aggregation method to generate reputation automatically based on the historical behaviours of clients, which is a more realistic approach for a real FL-based decentralised system implemented as a browser extension for clients.

\section{A Robust FL Method for Classifying Sensitive Content on the Web}\label{sec:FL}
In this section, we first show how to build an FL-based classifier for sensitive content. Then we design a reputation score for protecting against poisoning attacks. We analyse theoretically the combined FL/reputation-based solution and establish convergence and accuracy guarantees under common operating assumptions.

\subsection{FL Framework for Classifying Sensitive Content}
%\ci{The following paragraph reads a little bit weird. Maybe we need to rephrase it? We discuss about local models and a Global model. We can just say that clients provide updated parameters based on a common model that the serve aggregates to build the Global model or something simpler along these lines.}
Table~\ref{tab:notation} presents the notation that we use in the remainder of the paper. %Our goal of using FL for the sensitive classification task is to learn a global model $M_{global}$ for the server by aggregating local models of clients, which have the same structure as the global model. 
In FL, clients provide the server updated parameters from their local model, which the server aggregates to build the global model $M$. % global$. Here, we apply the model mentioned in Section~\ref{sec:performance} as the global model to classify five sensitive categories (Ethnicity, Health, Political Beliefs, Religion and Sexual Orientation) and one non-sensitive one.

Suppose we have $M$ clients participating in our classification training task and the dataset  $ \mathcal{D} = \bigcup_{i=1}^{M}\mathcal{D}_{i}$, where $\mathcal{D}_{i} \sim \mathcal{X}_{i} (\mu_{i},\,\sigma_{i}^{2})$ denotes the local data of client $i$ from non-independent and non-identically (Non-IID) distribution $\mathcal{X}_{i}$ with the mean $\mu_{i}$ and standard deviation $\sigma_{i}$. In our task, the clients' data is the textual content of URLs stripped of HTML tags. The objective function of FL,  $\mathcal{L}: \mathbb{R}^{d} \rightarrow \mathbb{R}$ which is the negative log likelihood loss in our task, can be described as \vspace{-1.5mm}
    $$\mathcal{L}(w) = \mathbb{E}_{D\sim \mathcal{X}}\left[l({w;\mathcal{D}}) \right]\vspace{-1.5mm}$$
where $l({w;\mathcal{D}})$ is the cost function of parameter $w \in \mathcal{W} \subseteq \mathbb{R}^{d}$. Here we assume $\mathcal{W}$ is a compact convex domain with diameter $d$. Therefore, the task becomes \vspace{-1.5mm}
    $$ w^{*} = \mathop{\arg\min}_{w \in \mathcal{W}} \mathcal{L}(w)\vspace{-1.5mm}$$
To find the optimal $w^{*}$, we employ Stochastic Gradient Descent (SGD) to optimise the objective function.

\begin{table}[t]
\caption{Notation}
\vspace{-2mm}
\adjustbox{max width=\columnwidth}{%
\centering
    \begin{tabular}{|l|l|}
\hline
Abbreviation  & Description                                                                      \\ \hline
$M$           & the total number of clients                                                      \\ \hline
$N$           & the number of parameters of global model                                         \\ \hline
$Q$           & the number of samples of each client                                             \\ \hline
$T$           & the total number of iterations                                                   \\ \hline
$w_{i,n}^{t}$ & the $n$-th parameter from client $i$ in $t$ iteration                            \\ \hline
$x_{i,n}^{t}$ & the  ranking of $w_{i,n}^{t}$ in  $w_{n}^{t}$                                    \\ \hline
$A_{n},B_{n}$ & the slope and intercept of repeated median linear regression                     \\ \hline
$e_{i,n}^{t}$ & the normalised residual of the $n$-th parameter from client $i$ in $t$ iteration \\ \hline
\end{tabular}
    }
    \label{tab:notation}
\end{table}

During the broadcast phase, the server broadcasts the classification task and training instructions to clients. Then, the clients apply the following standard pre-processing steps on the webpage content, that is, transformation of all letters in lowercase and the removal of stop words. Next, the clients extract the top one thousand features utilising the Term Frequency-Inverse Document Frequency (TF-IDF)~\cite{salton1988term} as in~\cite{matic2020identifying}.
% and send a dictionary with feature names and values to the server. After receiving all the dictionaries, the server ranks the features by aggregating their values, weighted by the percentage of each client's training sample size, and returns a list of the top 1k feature to the clients. Finally, clients prepare input for training based on this list from their own dataset. 
%Clients preprocess their text data using Term Frequency-Inverse Document Frequency (TF-IDF)~\cite{salton1988term} as in~\cite{matic2020identifying} with top 1k features.
%\ci{Expand a little bit on the classifier feature generation process. Paraphrase from the original IMC paper. Also include a list of the top features per category, similar to the old paper.} 
At iteration $t$, the client $i$ receives the current global model $M_{global}$ and then following the training instructions from server, trains the local model on its training data $\mathcal{D}_{i}$ and optimises $w_{i}^{t} = \mathop{\arg\min}_{w} \mathcal{L}_{i}(w_{i}^{t}) $ by using SGD: \vspace{-1.5mm}
$$w_{i}^{t} \leftarrow  w_{i}^{t-1} - r\frac{\partial \mathcal{L}_{i}(w_{i}^{t-1})}{\partial w}\vspace{-1.5mm}$$
where $\mathcal{L}_{i}(w):= \mathbb{E}_{\mathcal{D}_{i}\sim \mathcal{X}}[l(w;\mathcal{D}_{i})] =  \frac{1}{Q_{i}}\sum_{j=1}^{Q_{i}}l(w;\mathcal{D}_{i}^{j})$, $\mathcal{D}_{i}^{j}$ and $Q_{i}$ means the $j$-th sample and the number of samples of the client $i$ respectively, and $r$ is the learning rate.

In every iteration, after finishing the training process the clients send back their local updates to the server. Then, the server computes a new global model update by combining the local model updates via an aggregation method $\texttt{AGG}$ as follows: \vspace{-1.5mm}
$${w}^{t} = \texttt{AGG}\left (\left \{ {w}_{i}^{t} \right \}_{i=1}^{M} \right)\vspace{-1.5mm}$$ 
Here we utilise the basic aggregation method (FedAvg)~\cite{mcmahan2017communication}, which uses the fraction of each client's training sample size in total training samples as the average weights: \vspace{-1.5mm}
$${w}^{t} = \sum_{i=1}^{M}\frac{Q_{i}}{Q} {w}_{i}^{t}\vspace{-1.5mm}$$
%uses the fraction of each client's training sample size in total training samples as the weights to average local model updates of clients as the global model update. 
We introduce other robust aggregation methods in the next subsection. Subsequently, the server uses the global model update to renew the global model $M_{global}$.

% \ty{if we want to use the particular methods about  text classifier in FL, they must use SGD as the optimisation during training. all the method using SGD can be implemented here, we use a simple NN model in our method later. but the classifiers, e.g. Bayes in IMC-20, is not suitable here because they aren't trained in this way. about the plot, it would be a little overlap with Figure~\ref{fig:Loss}, and now we haven't described our aggregation method  which would be explained in Sect.3.2. but I would put here so you can decide if we need it or not.}
\begin{figure}[t]
    \centering
    \includegraphics[width=\columnwidth]{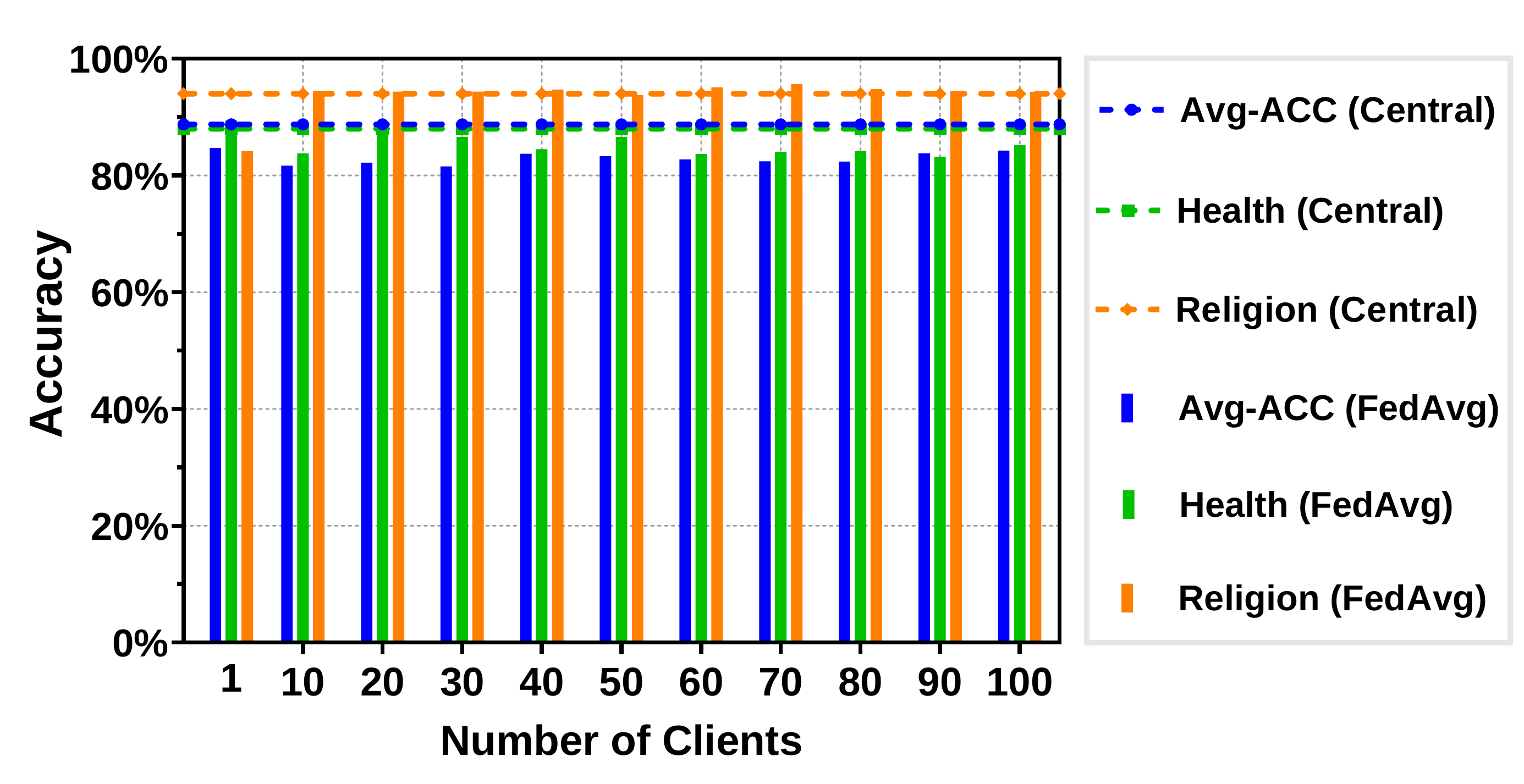}
    \vspace{-6mm}
    \caption{Accuracy of FL classifiers and centralised classifiers in Health, Religion and all category.}
    \label{fig:centralised}
\vspace{-1em}
\end{figure}
%\ci{suggestion to replace the following paragraph: Using the above FL-based framework we first evaluate how the number of users in the system affects the average accuracy of the classifier. The results for the two sensitive categories, Health and Religion as well as the overall average accuracy (Avg-ACC) are depicted in Figure~\ref{fig:centralised}. A first observation is that when a fixed size dataset is divided into multiple segments and distributed to more clients, the model's accuracy decreases since each client has less data for training. Compared to the centralised classifier, using the same data, the accuracy of the FL classifier is slightly lower, which is expected when the training is distributed to a larger number of clients. Overall, we observe that the average accuracy difference between the FL and the centralised classifier is 5.76\% which is acceptable. Looking at the different sensitive categories (Health and Religion) we see that the FL-based classifier achieves an accuracy very close to the corresponding one of the centralised classifier for these categories (on average 0.8533 vs. 0.88 and 0.9366 vs. 0.94, respectively).}
Using the above FL-based framework we first evaluate how the number of users in the system affects the average accuracy of the classifier. The results for the sensitive categories, Health and Religion, as well as the overall average accuracy (Avg-ACC) are depicted in Figure~\ref{fig:centralised}. A first observation is that when a fixed size dataset is divided into multiple segments and distributed to more clients, the model's accuracy decreases since each client has less data for training. Compared to the centralised classifier, using the same data, the accuracy of the FL classifier is slightly lower, which is expected when the training is distributed to a larger number of clients. Overall, we observe that the average accuracy difference between the FL and the centralised classifier is 5.76\%, and this remains steady as the number of clients grows. In addition, Looking at the different sensitive categories (Health and Religion), we see the FL-based classifier achieves an accuracy very close to the corresponding one of the centralised classifier for these categories (on average 0.8533 vs. 0.88 and 0.9366 vs. 0.94, respectively).

%Based on this FL framework for classifying sensitive content, %we apply the model mentioned in Section~\ref{sec:performance} as the global model $M_{global}$ and train it in a federated manner. 
%we show the initial result in Figure~\ref{fig:centralised} that the impact of increasing the total number of clients on average accuracy with a fixed text dataset (see details about this dataset in Section~\ref{sec:performance}). We show results on average accuracy as well as results for the Health and Religion sensitive categories as described in~\cite{matic2020identifying}. A first observation from these results is that when a fixed dataset is divided into multiple segments and distributed to more clients, the model's accuracy decreases. This is due to the fact that each client has less data for training. Compared to the centralised classifier using the same data, the FL classifier is slightly less accurate. Losing accuracy is normal when training is distributed to a larger number of clients. The average difference in average accuracy between FL classifier and centralised classifier is 5.76\% and %the gap is always less than 7\% and the overall loss of accuracy (3\%) \ci{we do not report loss in the figure.} throughout distribution is acceptable. Looking at the particular sensitive categories  of Health and Religion we see that the FL-based classifier achieves an accuracy very close to the corresponding one of the centralised classifier for these categories (on average 0.8533 vs. 0.88 and 0.9366 vs. 0.94, respectively). %\nla{Tianyue please add the missing numbers}.  

\subsection{A Reputation score for Thwarting Poisoning Attacks}

\begin{figure}[!bpt]
    \centering
    \includegraphics[width=0.5\textwidth,keepaspectratio]{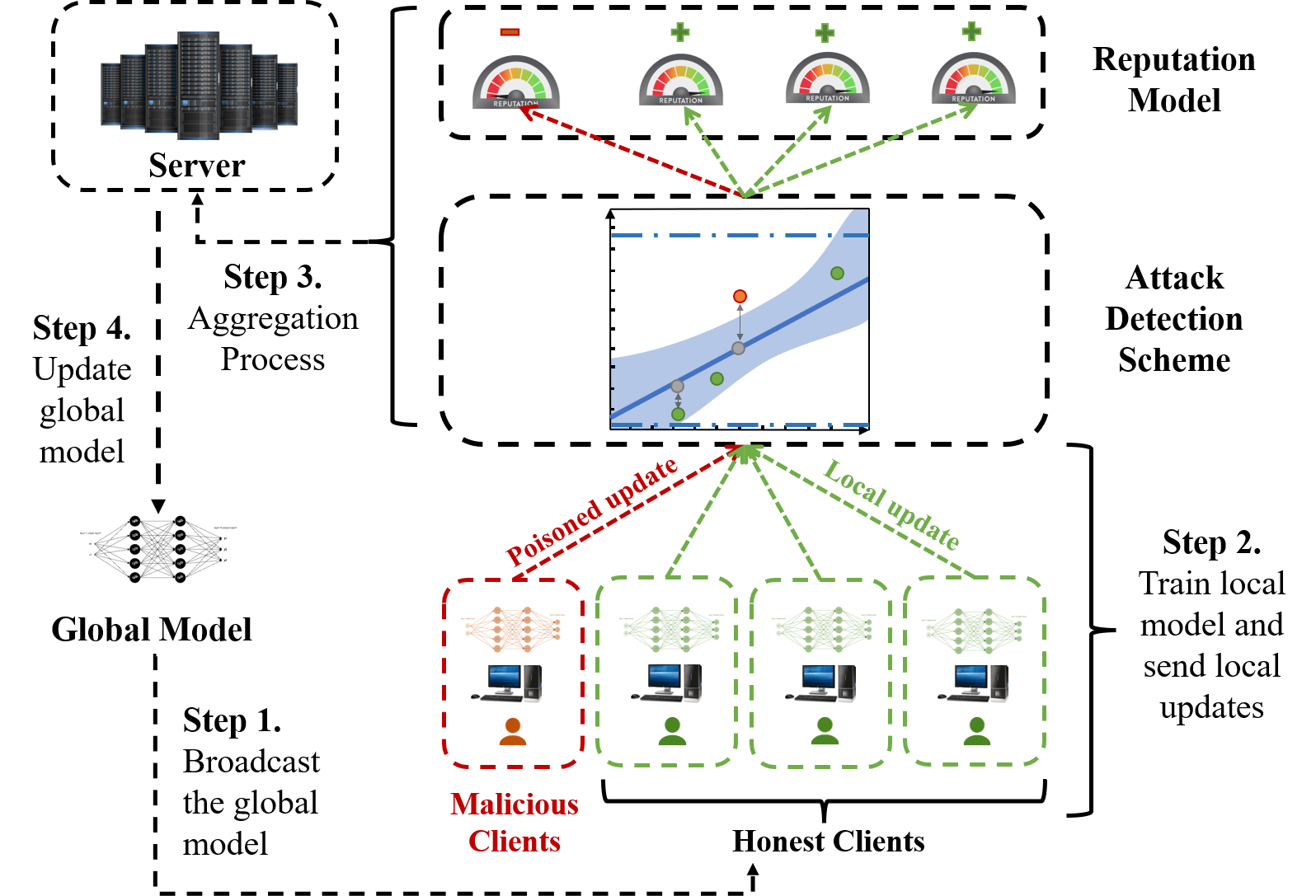}
    %\vspace{-7mm}
    \caption{Overview of reputation-based aggregation algorithm.}
    \label{fig:FL_architecture}
\vspace{-2em}
\end{figure}
%Whereas on the one hand, FL protects the privacy of clients by not exposing directly their local labels, on the other, it opens up the door to adversarial manipulations. For example, malicious clients may attempt to impede the process by sending misleading updates. To do this, malicious clients may intentionally ``poison'' their local training by using false labels to train it, and then share the resulting updates with other clients via the server. To address this problem, we  design an aggregation method for federated learning that achieves high robustness and accuracy against attacks. \nla{I think this has been repeated already several times. Maybe we can omit this intro paragraph to save space}

Figure~\ref{fig:FL_architecture} shows an overview of our reputation-based aggregation algorithm consisting of three components: the \emph{attack detection scheme}, the \emph{reputation model}, and the \emph{aggregation module}. The attack detection scheme re-scales and rectifies damaging updates received from clients. Then, the reputation model calculates each client's reputation based on their past detection results. Finally, the aggregation module computes the global model by averaging the updates of the clients using their reputation scores as weights. We detail each component in the following subsections.

\subsubsection{Attack Detection Scheme}\label{sub:attack detection scheme} 
Our attack detection scheme aims to reduce the impact of suspicious updates by identifying them and applying a rescaling algorithm. At every iteration, when model updates from clients arrive at the server, we apply Algorithm~\ref{al:rescale} there to rescale the range of values for those parameters in the updates.

This restriction on the value range aims not only to minimise the impact of  abnormal updates from attackers but also to limit the slope for the repeat median regression. Considering the $n$-th parameter in round $t$ from all the participants, we calculate the standard deviation $\sigma(w_{i,n}^{t})$ of this series. Then we sort them in ascending order and determine the range by subtracting the lowest value from the highest one. If the result is above the threshold $\varpi$, we rescale the highest and lowest value by deducting and adding its standard deviation respectively to further bound their range.

\begin{algorithm}[bpt]
   % \small
   \SetKwInOut{Input}{Input}
   \SetKwInOut{Output}{Output}
   
    \SetAlgoLined
     \Input{
     $\left \{ w_{i,n}^{t} \right \}$ $\leftarrow$ Local Model parameters in round $t$}
     
    \Output{$\left \{ w_{i,n}^{t} \right \}$ with range of value less than $\varpi$}

    \BlankLine
    \For {$n\leftarrow 1$ \KwTo $N$}{
      
      %Sort all $w_{i,n}^{t}$, $i = 1, \cdots,M$ in ascending order\;
    \tcp{Determine the maximum range}
       $Rm = \max{w_{i,n}^{t}} - \min{w_{i,n}^{t}} = w_{i,n}^{t,(Max)}-w_{i,n}^{t,(Min)}$
      \BlankLine
      \While{$Rm > \varpi$}{
      \tcp{Rescale range based on standard deviation.}
      $w_{i,n}^{t,(Max)} := w_{i,n}^{t,(Max)}- \sigma(w_{i,n}^{t}) $\;
      $w_{i,n}^{t,(Min)} := w_{i,n}^{t,(Min)}+ \sigma(w_{i,n}^{t}) $\;
      \tcp{Updated $Rm$.}
      $Rm = \max{w_{i,n}^{t}} - \min{w_{i,n}^{t}} = w_{i,n}^{t,(Max)} - w_{i,n}^{t,(Min)}$ 
      }
      }
\caption{Rescale($w$)
}
\label{al:rescale}
\end{algorithm}

Then, a robust regression~\cite{siegel1982robust} is carried out to identify outliers among the updates in the current round. Outlier detection is a well-established topic in statistics. Robust regression methods often handle outliers by using the median estimators. Median-based aggregation methods have a rich and longstanding history in the area of robust statistics~\cite{minsker2015geometric}. However, the methods developed by the traditional robust statistics can only withstand a small fraction of Byzantine clients, resulting in a low ``breakdown point''~\cite{lopuhaa1991breakdown}. Different from many other variations of the univariate median, the repeated median~\cite{siegel1982robust} is impervious to atypical points even when their percentage is nearly 50\%. The repeated median is defined as a modified U-statistic and the concept behind it is to utilise a succession of partial medians for computing approximation $\hat{\tau }$ of the parameter $\tau$: For $k\in \mathbb{N}$, the value of parameter $\tau (z_{1},\cdots,z_{k})$ is determines by subset of $k$ data points $z_{1},\cdots,z_{k}$.
\vspace{-1.5mm}
\begin{equation}
  \hat{\tau} = \underset{z_{1}}{\mathrm{median}}\left \{ \underset{z_{2}\notin \left \{ z_{1} \right \}}{\mathrm{median}} \left \{ \underset{z_{k}\notin \left \{z_{1},\cdots,z_{k-1} \right \}}{\mathrm{median}}\tau (z_{1},\cdots,z_{k}) \right \} \right \} 
\end{equation}
\vspace{-1.5mm}

In our case, the intercept $\hat{A}$ and slop $\hat{B}$ are estimated by repeated median as bellow: \vspace{-1.5mm}
\begin{equation}
  \hat{B}_{n} = \underset{i}{\mathrm{median}}\left \{ {\underset{i\neq j} {\mathrm{median}}\left \{ B_{n}(i,j)  \right \}} \right \} \vspace{-1.5mm}
\end{equation}
\vspace{-1.5mm}
\begin{equation}
  \hat{A}_{n} = \underset{i}{\mathrm{median}}\left \{ w_{i,n}-\hat{B}_{n}x_{i,n} \right \}\vspace{-1.5mm}
\end{equation}

\noindent where $B_{n}(i,j) = \frac{w_{j,n}-w_{i,n}}{x_{j,n}-x_{i,n}}$, $x_{i,n}$ represents the index of $w_{i,n}$ in $w_{n}$ which is sorted in ascending order.

Next, we employ the IRLS scheme~\cite{wilcox2011introduction} to generate each parameter's confidence score ${s}_{i,n}^{t}$ based on the normalised residual from repeated median regression, which is also utilised in a residual-based aggregation method~\cite{fu2019attack}: 
\begin{equation}\label{eq:parameter confidence}
\vspace{-1.5mm}
     {s}_{i,n}^{t} = \frac{\sqrt{1-diag(H_{n}^{t})}}{e_{i,n}^{t}}\Psi \left(\frac{e_{i,n}^{t}}{\sqrt{1-diag(H_{n}^{t})}}\right)
\vspace{-1.5mm}
\end{equation}
where confidence interval  $\Psi(x)$:
\begin{equation*}
\vspace{-1.5mm}
    \Psi(x) = \mathrm{max}\{-\lambda\sqrt{2/M},\mathrm{min}(\lambda\sqrt{2/M},x)\}
\end{equation*}
and the hat matrix $H_{n}^{t}$: \vspace{-1.5mm}
\begin{equation*}% for double superscript: R^{e^{Tx}}
    H_{n}^{t}=x_{n}^{t}(x_{n}^{t^{T}}x_{n}^{t})^{-1}x_{n}^{t^{T}}
\end{equation*}
 with 
$e_{i,n}^{t}=\frac{25(M-1)\left(w_{i,n}^{t}-\hat{B}_{n}x_{i,n}^{t}-\hat{A}_{n}\right)}{37(M+4)\underset{i}{\mathrm{median}}\left (|w_{i,n}^{t}-\hat{B}_{n}x_{i,n}^{t}-\hat{A}_{n}|\right )}$.
% and $x_{n}^{t}=[ x_{1,n}^{t},x_{2,n}^{t},\cdots,x_{M,n}^{t}]^{T}$.

The distance between the point and the robust line is described by the confidence score derived from the normalised residual, which can be used to evaluate if the point is anomalous. Following the computation of the parameter's confidence score, and in light of the fact that some attackers want to generate updates with abnormal magnitudes in order to boost the damage, a useful protection is to identify low confidence values based on a threshold $\delta$. Once the server recognises an update $\mathbf{w}_{i,n}^{t}$ of the client $i$ with confidence values less than $\delta$, rather than altering this update to the repeat median estimation, our technique replaces it with the median of $\mathbf{w}_{n}^{t}$, as follows: \vspace{-1.5mm}
\begin{equation}\label{eq:correction}
    {\mathbf{w}}_{i,n}^{t} =\begin{cases}
{\mathbf{w}}_{i,n}^{t}  & \text{ if } {s}_{i,n}^{t}>\delta \\ 
\underset{i}{\mathrm{median}}\left \{{\mathbf{w}}_{i,n}^{t} \right \} &  \text{ if }{s}_{i,n}^{t}\leq \delta
\end{cases}
\vspace{-1.5mm}
\end{equation}
With the above, not only we bound the range of updates, but also improve the aggregation by introducing a robustness estimator.

\subsubsection{Reputation Model}

During the aggregation phase in FL, we use a subjective logic model to produce client reputation scores. The subjective logic model is a subset of probabilistic logic that depicts probability values of belief and disbelief as degrees of uncertainty~\cite{josang2016subjective}. 
%, which is an useful model for explaining about the  behaviour of percentages and proportions.
In the subjective logic model, reputation score $R_{i}^{t}$ for client $i$ in $t$ iteration correlates to a subjective belief in the dependability of the client's behaviour~\cite{kang2019incentive}, as measured by the belief metric opinion \(\tau_{i}^{t}\)~\cite{josang2006trust}. An opinion is comprised of three elements: belief $b_{i}^{t}$, disbelief $d_{i}^{t}$ and uncertainty $u_{i}^{t}$, with restrictions that
$b_{i}^{t}+d_{i}^{t}+u_{i}^{t}=1$ and $b_{i}^{t},d_{i}^{t},u_{i}^{t}\in \left [ 0,1 \right ]$. The reputation score may be calculated as the expected value of an opinion $E(\tau_{i}^{t})$ which can be regarded as the degree of trustworthiness in client $i$. As a result, the value of the client's reputation is defined as follows: \vspace{-1.5mm}
\begin{equation}\label{eq:opinion expectation}
R_{i}^{t} = E(\tau_{i}^{t}) = b_{i}^{t} + au_{i}^{t}
\vspace{-1.5mm}
\end{equation}
where $a\in \left [ 0,1 \right ] $ denotes the prior probability in the absence of belief, which reflects the fraction of uncertainty that may be converted to belief. 
On the other side, distinct observations determined by the rectification phase in our Algorithm~\ref{al:aggregtaion} are used to count belief, disbelief, and uncertainty opinions. The positive observation denoted by $P_{i}^{t}$ indicates that the update $\mathbf{w}_{i,n}^{t}$ is accepted (${s}_{i,n}^{t}>\delta$), whereas a negative observation denoted by $N_{i}^{t}$ indicates that the update is rejected (${s}_{i,n}^{t}\leq\delta$). As a consequence, the positive observations boost the client's reputation, and vice versa. To penalise the negative observations from the unreliable updates, a higher weight $\eta$ is assigned to negative observations than the weight $\kappa$ to positive observations with constrain $\eta + \kappa = 1$. Therefore, in Beta distribution below: \vspace{-1.5mm}
\begin{equation} \label{eq:beta distribution}
     Beta(p|\alpha ,\beta)= \frac{\Gamma (\alpha+\beta)}{\Gamma (\alpha)\Gamma (\beta)}x^{\alpha-1}(1-x)^{\beta-1}
\vspace{-1.5mm}
\end{equation}
with the constraints $0\leq x\leq 1$, parameters $\alpha > 0, \beta> 0$, and $x \neq 0$ if $\alpha < 1$ and $x \neq 1$ if $\beta < 1$. The parameters $\alpha$ and $\beta$ that represent positive and negative observations respectively, can be expressed as below \vspace{-1.5mm}
\begin{equation}\label{eq:alphabeta}
\begin{cases}
 & \alpha= \kappa P_{i}^{t} + Wa \\ 
 & \beta = \eta N_{i}^{t} + W(1-a)\\
\end{cases}
\vspace{-1.5mm}
\end{equation}
where $W$ is the non-information prior weight and the default value is 2~\cite{josang2016subjective}. 

As a consequence, the expected value of Beta distribution, which also stands for reputation value, can be calculated as follows: \vspace{-1.5mm}
\begin{equation}\label{eq:beta expectation}
    E(Beta(p|\alpha ,\beta)) =  \frac{\alpha}{\alpha+\beta} = \frac{\kappa P_{i}^{t} + Wa}{\kappa P_{i}^{t} + \eta N_{i}^{t} + W} = R_{i}^{t}
\vspace{-1.5mm}
\end{equation}

\noindent Based on \eqref{eq:opinion expectation} and \eqref{eq:beta expectation}, we can derive \vspace{-1.5mm}
\begin{equation}\label{eq:belief metric}
\left\{\begin{matrix}
b_{i}^{t} = \frac{\kappa P_{i}^{t}}{ \kappa P_{i}^{t} + \eta N_{i}^{t} + W}\\ 
d_{i}^{t} = \frac{\eta N_{i}^{t}}{ \kappa P_{i}^{t} + \eta N_{i}^{t} + W}\\ 
u_{i}^{t} = \frac{W}{ \kappa P_{i}^{t} + \eta N_{i}^{t} + W}\\
\end{matrix}\right.
\vspace{-1.5mm}
\end{equation}

In addition, in order to take the client's historical reputation values in previous rounds into consideration, a time decay mechanism is included to lower the relevance of past performances without disregarding their influence. In other words, the reputation value from the most recent iteration contributes the most to the reputation model. We use exponential time decay in our model, as shown below: \vspace{-1.5mm}
\begin{equation}
    \theta_{j,t} = \exp(-c(t-j))
\vspace{-1.5mm}
\end{equation}
where $\exists c > 0$, $j \in \left [\tilde{s},t \right ]$, $\tilde{s}=\mathrm{max}\left (t-s,0 \right)$.
We include a sliding window with a window length $s$ that allows us to get a reputation for a certain time interval rather than the entire training procedure. We remove expired tuples with timestamps outside the window period during computation since they cannot provide meaningful information for the reputation. 
% Hence, the opinion can be expressed as
% \begin{equation} \label{eq:time belife}
% \begin{cases}
% % \left\{\begin{matrix}
% \tilde{b}_{i}^{t}=\frac{\sum_{j=t-s}^{t}\theta_{j,t}b_{i}^{j}}{\sum_{j=t-s}^{t}\theta_{j,t}}\\ 
% \tilde{d}_{i}^{t}=\frac{\sum_{j=t-s}^{t}\theta_{j,t}d_{i}^{j}}{\sum_{j=t-s}^{t}\theta_{j,t}}\\
% \tilde{u}_{i}^{t}=\frac{\sum_{j=t-s}^{t}\theta_{j,t}u_{i}^{j}}{\sum_{j=t-s}^{t}\theta_{j,t}}\\
% \end{cases}
% % \end{matrix}\right.
% \end{equation}
Hence, the final reputation score $\tilde{R}_{i}^{t}$ can be expressed as: %\vspace{-1.5mm}
\begin{equation} \label{eq:final reputation}
    \tilde{R}_{i}^{t} = \frac{\sum_{j=\tilde{s}}^{t}\theta_{j,t}R_{i}^{j}}{\sum_{j=\tilde{s}}^{t}\theta_{j,t}}
%\vspace{-1.5mm}
\end{equation}

To demonstrate how the reputation model evolves, we consider four scenarios where each client:
(i) only attacks once at the same iteration,
(ii) attacks continuously after launching an attack at the same iteration,
(iii) only attacks once at different iteration,
(iv) attacks continuously after launching an attack at different iteration.
Here, clients conduct attacks as described in Section~\ref{sec:threat model} by utilising polluted data while training the local model, whereas the server uses our attack detection mechanism to identify these attacks. 

\begin{figure}[!t]
    \centering
    \includegraphics[width=\columnwidth]{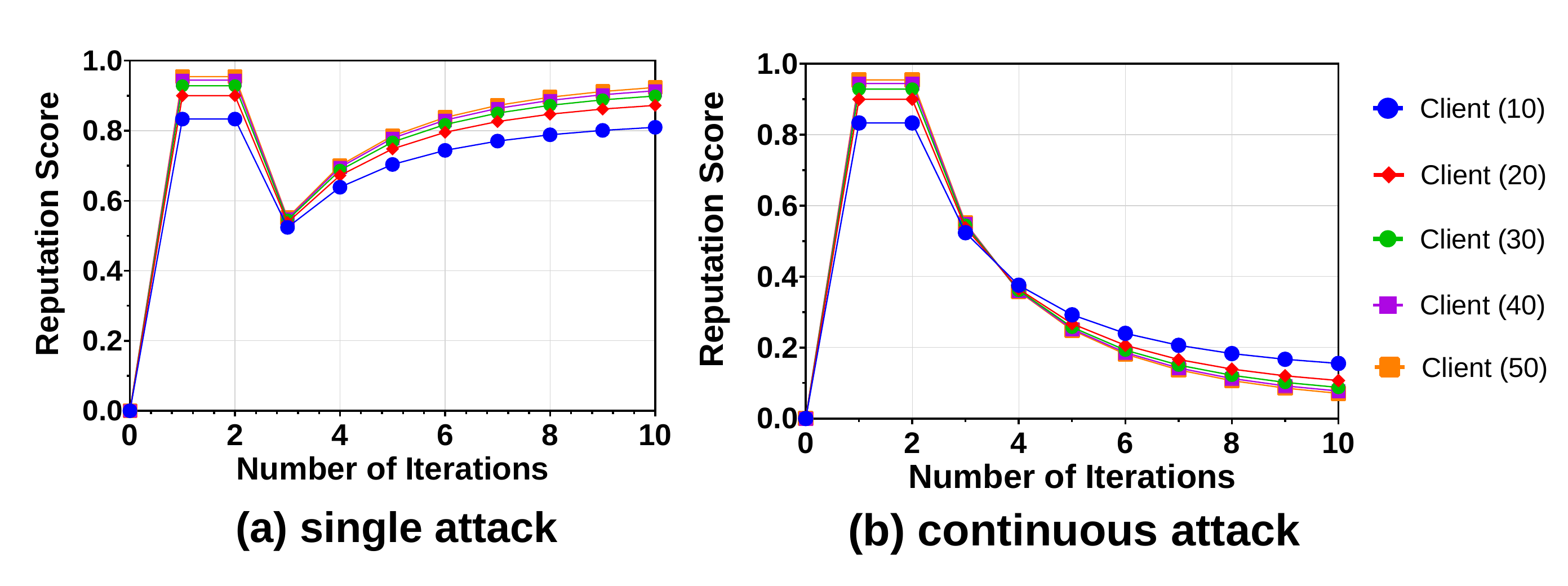}
    %\vspace{-6mm}
    \caption{The decay of reputation score in Client ($X$) with $X$ model parameters when they (a) attack once at 3rd iteration and (b) attack continuously at and after the 3rd iteration.}
    \label{fig:fix_rep}
\vspace{-1em}
\end{figure}

Figure~\ref{fig:fix_rep} displays the first two scenarios (i)-Figure~\ref{fig:fix_rep}a and (ii)-Figure~\ref{fig:fix_rep}b, respectively with Client $X$, who has $X$ parameters in their local models, under single and continuous attack. In Figure~\ref{fig:fix_rep}a, all of the clients only attack once at the third iteration. When they start attacking, their reputation score plummets dramatically. In both scenarios, we observe the client who has more parameters has a larger relative decline in reputation score. This is also compatible with Corollary~\ref{co:corollary1} in the Section~\ref{sec:theory}, that is, increasing the number of parameters $N$ in the global model results in a lower error rate. %\nla{Shouldn't we remind what the parameters are here? Tokens from TF-IDF and give some examples, e.g., for Health?} \ty{parameters here are model parameters, which is the weights of neural network, depending on the structure of the global model we choose.}

Figure~\ref{fig:varied_rep} shows the last two scenarios (iii) and (iv) respectively with clients, who have 20 parameters in their local models, under single and continuous attack.
% The deterioration of 10 users' reputation scores is seen in Figure., when they launch an attack.
In Figure~\ref{fig:varied_rep}a, Client $X$ only launches an attack at $X$ iteration. We observe that only one attack would lead to at least a 25.11\% relative decrease in reputation score. In Figure~\ref{fig:varied_rep}b, Client $X$ launches an attack at $X$ iteration and keeps attacking in the following iterations. We observe in the end that 80\% of their reputation scores are below 0.5, which is approximately half of the reputation score of honest clients, implying that the damage that they can inflict throughout the aggregation process is considerably decreased.

\begin{figure}[!t]
    \centering
    \includegraphics[width=\columnwidth]{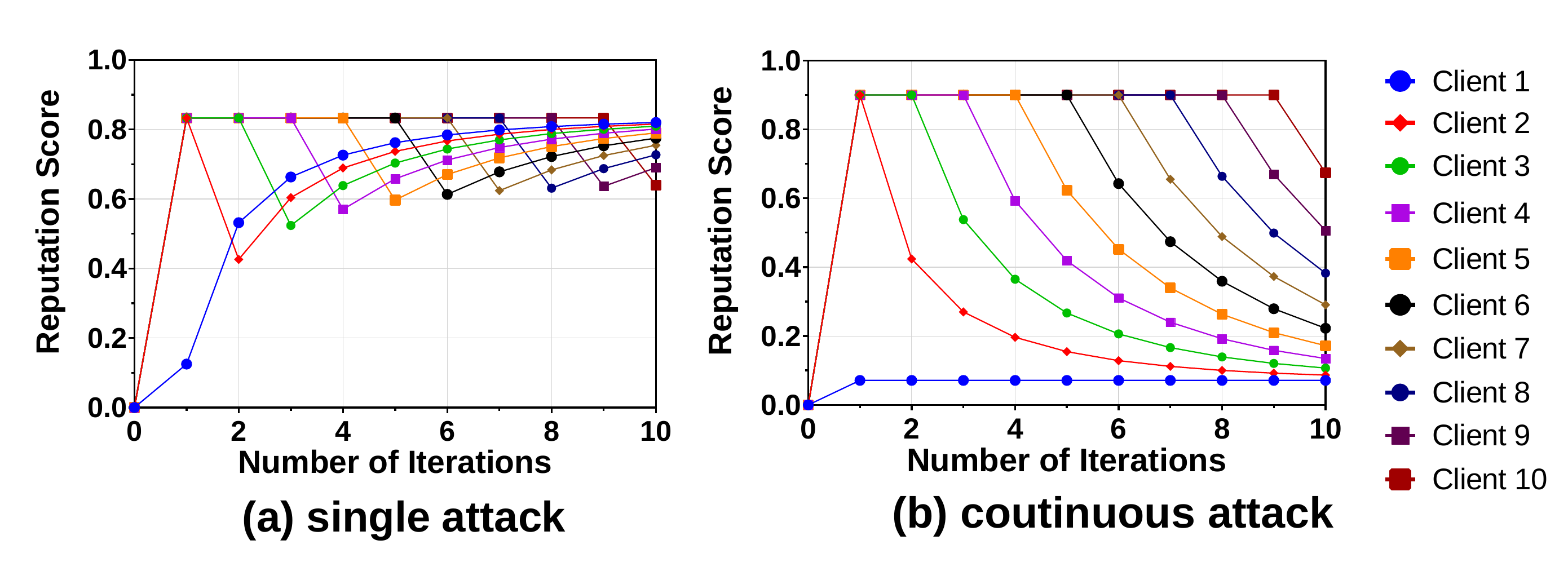}
    %\vspace{-6mm}
    \caption{The decay of reputation score in Client $X$ with same model parameters when they (a) attack once at $X$ iteration and (b) attack continuously after starting to attack at $X$ iteration.}
    \label{fig:varied_rep}
\vspace{-2em}
\end{figure}

In addition, we consider a scenario in which an attacker spreads out the poisoning over a longer time duration, while using a higher number of model parameters. Figure~\ref{fig:rep_loop} (left) depicts an attack over 40 iterations under different parameter sizes. Figure~\ref{fig:rep_loop} (right) depicts an attack with 1 million parameters repeating every 50 to 80 iterations.
These figures show that even if attackers spread 
our poisoning over multiple iterations and then try to recover their reputation score by acting benignly, our detection scheme can still identify them. This is because our attack detection and reputation schemes work in sequence. The attack detection scheme detects malicious updates without considering any reputation scores and rectifies them to mitigate damage. Then, the reputation scheme modifies the reputation scores based on the detection results. 
Also, attackers that employ a higher number of model parameters suffer a slightly higher reduction of reputation, which is consistent with Corollary~\ref{co:corollary1}.
\begin{figure}[!t]
    \centering
    \includegraphics[width=\columnwidth]{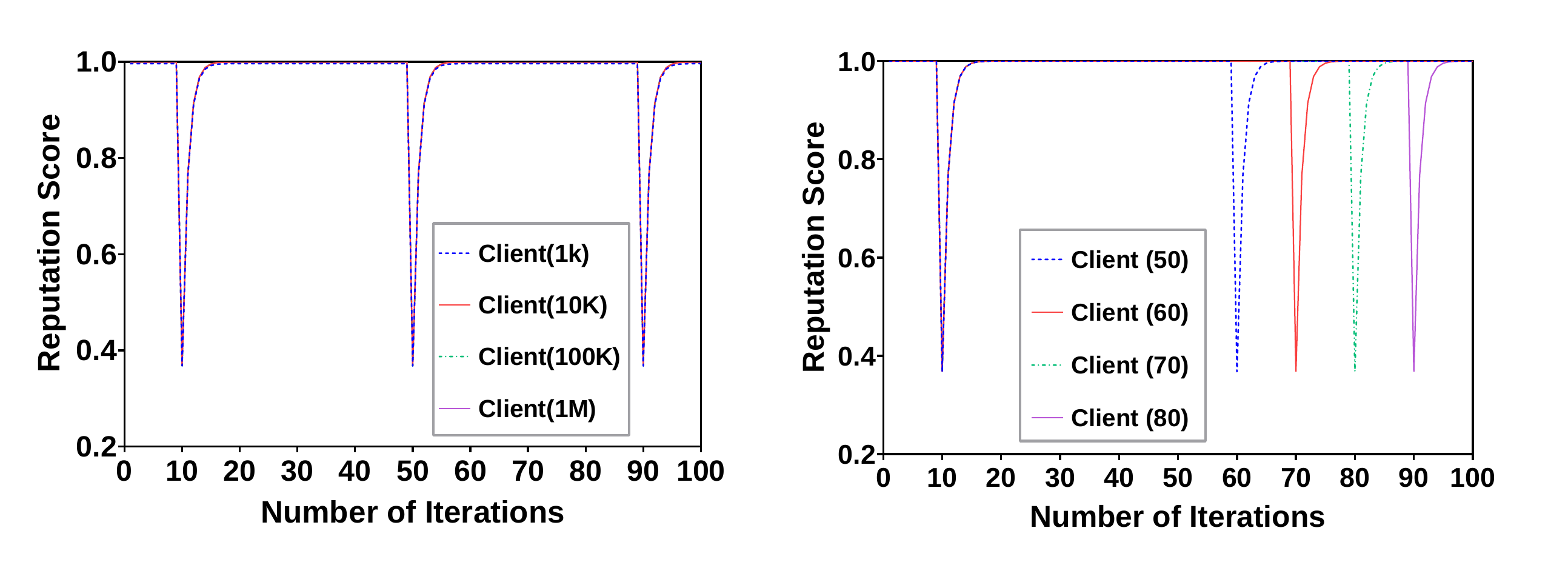}
    %\vspace{-6mm}
    \caption{The decay of reputation score in (left) Client with $X$ model parameters when they attack at 10, 50 and 90 iteration; (right) Client $X$ with 1 million parameters when they attack at 10 and $10+X$ iteration.}
    \label{fig:rep_loop}
    \vspace{-1em}
\end{figure}

\subsubsection{Aggregation Algorithm}
Algorithm~\ref{al:aggregtaion} explains our aggregation method based on the attack detection scheme and subjective logic reputation model. First, the server sends all clients the pre-trained global model with initial parameters. Then, using their own data samples, clients train the global model locally and send the trained parameters back to the server. At this point, the server executes the attack detection scheme. In round $t$, if the $n$-th update parameter $\mathbf{w}_{i,n}^{t}$ from the client $i$ has been rectified by the attack detection scheme in Section~\ref{sub:attack detection scheme} to the median value, the server regards it as a negative observation, whereas no rectification represents a positive observation. Then, the server punishes the negative observation by reducing the corresponding client's reputation. Both types of observations are accumulated through all the $N$ parameters of client $i$ to obtain the reputation value $\tilde{R}_{i}^{t}$ in $t$ round for client $i$ so as to all the other clients. The server would conduct Min-Max normalisation to obtain $\bar{R}^{t}$ after receiving the reputation values $\tilde{R}^{t}$ of all the clients in $t$ round. 

\begin{algorithm}[t]
    %\small
   \SetKwInOut{Input}{Input}
   \SetKwInOut{Output}{Output}
   \SetKwInOut{Server}{Server}
   \SetKwInOut{Client}{Client}
   \SetKwFunction{Rescale}{Rescale}
   \SetKwFunction{Norm}{Norm}
   
   \SetAlgoLined
   \Server{}
    \Input{
    $w^{0}$ $\leftarrow$ Pretrained\ Model\\ %$initialization$\\
    $\kappa$,$\eta$,$a$,$W$,$c$,$s$ $\leftarrow$ Reputation parameters}
    %$\tilde{R}^{0}\leftarrow$ Reputation\ score $initialization$}
    \Output{Global model $M_{global}$ with $w^{T}$}
     %\(y_{}^{0}\leftarrow pretrained\ model \left \{ initialization \right \}\) \;
    \BlankLine
    \For{Iteration $t\leftarrow 1$ \KwTo $T$}
    {
    \tcp{Broadcast global model to clients}
    \textbf{send}$(w^{t-1})$\;
    \tcp{Wait until all updates arrive} \textbf{receive}$(w^{t})$\;
    \tcp{Rescale parameters by Algorithm~\ref{al:rescale}}
    $\bar{w}^{t} \leftarrow $ \Rescale{$w^{t}$}\;
    \For {$n\leftarrow 1$ \KwTo $N$}{
      \For {$i\leftarrow 1$ \KwTo $M$}
      {
           \tcp{Compute parameter confidence}
     ${s}_{i,n}^{t}= Eq$~\ref{eq:parameter confidence}($\bar{w}_{i,n}^{t}$)\;
        %update  \(\left \{ v_{i}^{t}: i \in \left [ m \right ] \right \}\) \;
    % Set threshold $\delta$ on parameter confidence\;
    % Replace ${s}_{i,n}^{t}$ with ${s}_{i,n}^{t}\mathbbm{1}({s}_{i,n}^{t}>\delta)$\;
          \tcp{Rectify abnormal parameters} 
          $w_{i,n}^{t}:= Eq$~\ref{eq:correction}  (${s}_{i,n}^{t},\delta$);\\
    % and $\tilde{R}_{i}^{t-1}$ ;\\
    \textbf{record }$(P_{i}^{t},N_{i}^{t})$\;
    %Reweigh %$W_{i}=\tilde{R}_{i}^{t}\cdot\sum_{n=1}^{N}{s}_{i,n}^{t}\sigma({s}_{i,n}^{t}) $
    }
    %\(x^{t}\leftarrow x^{t-1}-\gamma \sum_{i=1}^{M}\frac{\tilde{W_{i}}}{\tilde{W}}v_{i}^{t}\)}
    }
    %$y_{n}^{t} = \sum_{i=1}^{M}\frac{W_{i}}{\sum_{i=1}^{M}W_{i}}y_{i,n}^{t}$}
    \For {$i\leftarrow 1$ \KwTo $M$}{
    \tcp{Calculate reputation score} $\tilde{R}_{i}^{t} = Eq$~\ref{eq:final reputation}($P_{i}^{t}$, $N_{i}^{t}$, $\kappa$, $\eta$, $a$, $W$, $c$, $s$)\; 
    }
    \tcp{Normalisation}
    $\bar{R}^{t} \leftarrow $ \Norm{$\tilde{R}^{t}$}\;

    \For {$n\leftarrow 1$ \KwTo $N$}{
    \tcp{Update the parameters} 
    $w_{n}^{t} := \sum_{i=1}^{M}\frac{\bar{R}_{i}^{t}}{\sum_{i=1}^{M}\bar{R}_{i}^{t}}w_{i,n}^{t}$\;
    }
    \tcp{Obtain parameters for global model}
    $w^{t} :=  \left [ w_{1}^{t},\cdots,w_{n}^{t} \right ]$\;
    }
    
    \BlankLine
    \Client{}
    \BlankLine
    \setcounter{AlgoLine}{0}
    \ForPar{Client $i\leftarrow 1$ \KwTo $M$}{
     \textbf{receive}$(w^{t-1})$\;
     \tcp{Train local model}
     $w_{i}^{t} \leftarrow  w_{i}^{t-1} - r\frac{\partial \ell_{i}(w_{i}^{t-1})}{\partial w}$\;
      \textbf{send}($w_{i}^{t}$)\;
     % the gradient \(v_{i}^{t}=\bigtriangledown F_{i}^{t}(x^{t-1})\) to the server
     }
\caption{Aggregation Algorithm}
\label{al:aggregtaion}
\end{algorithm}

After the server gets correction updates and the normalised reputation of each client, it aggregates the updates using average weighted reputation as the weights to get our global model updates for the current iteration. In this way, even over many training rounds, the attackers are still incapable of shifting parameters notably from the target direction and this ensures the quality of the resulting global model as will be demonstrated experimentally and analytically next.

\subsection{Theoretical Guarantees}\label{sec:theory}

%In this section, 
We prove the convergence of our reputation-based aggregation method. %\nla{Our major result is Theorem X that states that [some qualitative result like: convergence is guaranteed in bounded, or polynomial, or whatever amount of time]}
Our major results are Theorem~\ref{th:theorem1} and Corollary~\ref{co:corollary1}, which state that convergence is guaranteed in bounded time. Regarding the performance of our algorithm in terms of metric average accuracy and convergence, 
%, as will be shown later, 
we show that it is consistent with our theoretical analysis. We start by stating our assumptions, which are standard and common for such types of results, and per recent works such as~\cite{yin2018byzantine, xie2019zeno}.

\begin{assumption}[Smoothness]\label{as:assumption1}
The loss functions  are $L$-smooth, which means they are continuously differentiable and their gradients are Lipschitz-continuous with Lipschitz constant $L>0$, whereas: \vspace{-1.5mm}
\begin{gather*}
\forall i \in N,\, \forall \mathbf{w}_{1},\mathbf{w}_{2} \in \mathbf{R}^{d}\\
\| \nabla{\mathcal{L}(\mathbf{w}_{1}))}-\nabla{\mathcal{L}(\mathbf{w}_{2}))} \|_{2} \leq L \left \|\mathbf{w}_{1}-\mathbf{w}_{2}\right \|_{2}\\
\left \| \nabla{\ell(\mathbf{w}_1;\mathcal{D})}-\nabla{\ell(\mathbf{w}_2;\mathcal{D})}\right \|_{2} \leq L \| \mathbf{w}_{1}-\mathbf{w}_{2} \|_{2}
\end{gather*}
% The loss function of client $l(w_{i},D)$ as well as server $\mathcal{L}(w)$ are L-smooth on $\mathbb{R}^{d}$, which means they are continuously differentiable and their gradient are Lipschitz-continuous with Lipschitz constant $L >0 $, whereas: 
% $\forall w_{1},w_{2} \in \mathcal{W} \in \mathbb{R}^{d}$,
% $\left \| \nabla{l(w_1;\mathcal{D})}-\nabla{l(w_2;\mathcal{D})}\right \|_{2} \leq L \| w_1-w_2 \|_{2}$
% and 
% $\| \nabla{\mathcal{L}(w_1))}-\nabla{\mathcal{L}(w_2))} \|_{2} \leq L \left \| w_1-w_2\right \|_{2}$
\end{assumption}

\begin{assumption}[Bounded Gradient]\label{as:assumption2}
The expected square norm of gradients $\mathcal{w}$ is bounded: \vspace{-1.5mm}
\begin{gather*}
    \forall \mathbf{w} \in \mathbf{R}^{d},  \exists\mathcal{G}_{\mathbf{w}} < \infty, \mathbb{E}\left \| \nabla{\ell}(\mathbf{w};\mathcal{D}) \right \|_{2}^{2}\leq \mathcal{G}_{\mathbf{w}}
\end{gather*}
% The gradient $w$ is bounded. $\forall w \in \mathcal{W}$, $ \exists\mathcal{G}_{w} < \infty$, $\left \| \nabla{l}(w;\mathcal{D}) \right \|\leq \mathcal{G}_{w}$
\end{assumption}

\begin{assumption}[Bounded Variance]\label{as:assumption3}
The variance of gradients $\mathbf{w}$ is bounded: \vspace{-1.5mm}
\begin{gather*}
\forall \mathbf{w} \in \mathbf{R}^{d},  \exists\mathcal{V}_{\mathbf{w}} < \infty, \mathbb{E}\left \| \nabla{\ell}(\mathbf{w};\mathcal{D})-\mathbb{E}(\nabla{\ell}(\mathbf{w};\mathcal{D}) \right \|_{2} ^2 \leq \mathcal{V}_{\mathbf{w}} 
\end{gather*}
% The variance of gradient $w$ is bounded. $\forall w \in \mathcal{W}$, $ \exists\mathcal{V}_{w} < \infty$, $\mathbb{E}\left \| \nabla{l}(w;\mathcal{D})-\mathbb{E}(\nabla{l}(w;\mathcal{D}) \right \| ^2 \leq \mathcal{V}_{w}$
\end{assumption}

\begin{assumption}[Convexity]\label{as:assumption4} 
The loss function $\mathcal{L}(\mathcal{w})$  are $\mu$-strongly convex: \vspace{-1.5mm}
\begin{gather*}
 \exists \mu > 0,  
 \forall \mathbf{w}_{1},\mathbf{w}_{2} \in \mathbf{R}^{d}, \nabla{\mathcal{L}(\mathbf{w}^{*})} = 0\\
 \mathcal L(\mathbf{w}_{1})- \mathcal L(\mathbf{w}_{2}) \geq \left \langle \nabla{\mathcal L(\mathbf{w}_{2})}, \mathbf{w}_{1}-\mathbf{w}_{2}\right \rangle + \frac{\mu}{2}\left \| \mathbf{w}_{1}-\mathbf{w}_{2} \right \|_{2}^{2}
\end{gather*}
% $\mathcal{L}(\mathbf{w})$ is a $\mu$-strongly convex. $\exists \mu > 0$ such that $\forall \mathbf{w}_{1}, \mathbf{w}_{2} \in \mathcal{W}$, $\mathcal L(w_{1})- \mathcal L(\mathbf{w}_{2}) \geq \left \langle \nabla{\mathcal L(\mathbf{w}_{2})}, \mathbf{w}_{1}-\mathbf{w}_{2}\right \rangle + \frac{\mu}{2}\left \| \mathbf{w}_{1}-\mathbf{w}_{2} \right \|_{2}^{2}$, and $\nabla{\mathcal{L}(\mathbf{w}^{*})} = 0$ 
\end{assumption}
\vspace{-3mm}
Suppose the percentage of attackers in the whole clients is $p$, and all the clients in the system participant every training iteration. $r$ is the learning rate($r < \frac{1}{L}$) and $\hat{Q} = \mathrm{max}\left \{Q_{i}  \right \}_{i=1}^{M}$. $\forall \mathbf{w}\in  \mathcal{W}$, we denote \vspace{-1.5mm}
$$\mathbf{m}_{i}(\mathbf{w}^{t}) = \begin{cases}
* & \text{ if } i \in {\mbox{malicious clients}} \\ 
\nabla{l_{i}(\mathbf{w}^{t};\mathcal{D})} & \text{ if } i \in {\mbox{honest clients}}
\end{cases}\vspace{-1.5mm}$$
where $*$ stands for an arbitrary value from the malicious clients. \vspace{-1.5mm}
$$\mathbf{m}(\mathbf w^{t}) = \sum_{i=1}^{M}\bar{R}_{i}\mathbf m_{i}(\mathbf{w}^{t})$$ \vspace{-1.5mm}
$$s.t.\; \bar{R}_{i} = \frac{\tilde{R_{i}^{t}}}{\sum_{i=1}^{M}\tilde{R_{i}^{t}}},\;\sum_{i=1}^{M}\bar{R_{i}} = 1,\; \bar{R_{i}} \in (0,1) \vspace{-1.5mm} $$

Consider the assumptions above and lemmas presented in Appendix~\ref{sec:proofs}, we have 
\begin{theorem}
\label{th:theorem1}
Under Assumptions \ref{as:assumption1}, \ref{as:assumption2}, \ref{as:assumption3} and~\ref{as:assumption4}, $\exists \epsilon > 0$ that: \vspace{-1.5mm}
\begin{equation}
\label{ineq:probablity}
   \sqrt{\frac{d\log(1+\hat{Q}MLD_{\epsilon})}{M(1-p)}}+C\frac{\mathcal{G}_{\mathbf{w}}}{\sqrt{\hat{Q}}}+ p\leq \frac{1}{2} - \epsilon  
\end{equation} 
After $t$ rounds, Algorithm~\ref{al:aggregtaion} converges with probability at least $1-\xi \in \left [1-\frac{4d}{\left ( 1+\hat{Q}ML\upsilon \right)^{d}},1\right )$ as \vspace{-2mm}
\begin{align}
     \left \| \mathbf{w}^{t} - \mathbf{w}^{*} \right \|_{2} \leq & \left ( 1- Lr \right )^{t}\left \| \mathbf{w}^{0} - \mathbf{w}^{*} \right \|_{2}+ \frac{\sqrt{N}}{L}\Delta_{1} + \frac{1}{L}\Delta_{2}
\end{align}\vspace{-2mm}
where %\ci{the second equation overlaps with the right column even after using \textbackslash small. Is there any other solution to fix it?}
\vspace{-1.5mm}
$$\Delta_{1} = \frac{M\left(\varpi(M-1)+\frac{2E}{\sqrt{M}\delta}\right)}{\frac{Wa(M-1)(\kappa N + W)}{(\eta N + W)(\kappa N + Wa)}+1} \vspace{-1.5mm} $$

{\small $$\Delta_{2} = 2\sqrt{2}\frac{1}{M\hat{Q}} + \sqrt{\frac{2}{\hat{Q}}} D_{\epsilon}V_{w} \left( \sqrt{\frac{d\log(1+\hat{Q}ML\upsilon)}{M(1-p)}}+C\frac{\mathcal{G}_{w}}{\sqrt{\hat{Q}}}+p \right) $$} \vspace{-1.5mm}
$$D_{\epsilon} := \sqrt{2\pi}\exp\left ( \frac{1}{2}\left (\Phi (1-\epsilon ) \right )^{2} \right) \vspace{-1.5mm}  $$
with $\Phi \left ( \cdot  \right )$ being the cumulative distribution function of Wald distribution.
\end{theorem}
%the corollary below  \todo{this sentence feels sort of disconnected from the next corollary}
\begin{corollary}
\label{co:corollary1} Continuing with Theorem~\ref{th:theorem1},
when the iterations satisfy $t \geq \frac{1}{Lr}\log \left ( \frac{L}{\sqrt{N}\Delta_{1}+\Delta_{2}}\left \| \mathbf{w}^{0} - \mathbf{w}^{*} \right \|_{2} \right )$, $\exists \xi \in  \left (0, \frac{4d}{\left ( 1+\hat{Q}ML\upsilon \right)^{d}} \right]$, we have: \vspace{-1.5mm}
{\small $$\mathbb{P}\left (\left \| \mathbf{w}^{t}-\mathbf{w}^{*} \right \|_{2} \leq \frac{2\sqrt{N}}{L}\Delta_{1} + \frac{2}{L} \Delta_{2} \right )\geq 1-\xi $$}
\end{corollary}

\begin{remark}
\label{re:remark1}
Due to \vspace{-1.5mm}
%\agr{As discussed with Nikos, remarks are not clear, and as I said, as a reviewer, I stopped reading here. At least a small descriptive text to support equalities (not enumerated as such in latex btw) is needed in Remark 1 (e.g., Due to the sum of the two error rates for the two terms that equalities X and Y represent respectively with $\Delta_{1}$ and $\Delta_{2}$ ..., we achieve a total error rate Z based on Corollary 3.2). What a reviewer should read this for exactly? it is a proof (to appendix) to show empirical result is consistent with theory only.}
{\small $$\Delta_{1}:= \mathcal{O} \left(\frac{\varpi}{a\kappa W N} + \frac{1}{\kappa N}+ \frac{1}{\sqrt{M N}\delta} \right) $$} \vspace{-1.5mm}
and
{\small $$\Delta_{2}:= \mathcal{O} \left( \frac{1}{\hat{Q}} + \frac{p}{\sqrt{\hat{Q}}}+\frac{1}{\sqrt{\hat{Q}M}} \right)$$} \vspace{-1.5mm}
Based on Corollary~\ref{co:corollary1}, we achieve an error rate: \vspace{-1mm}
{\small $$\mathcal{O} \left(\frac{\varpi}{a\kappa W \sqrt{N}} + \frac{1}{\kappa \sqrt{N}} + \frac{1}{\sqrt{M}\delta}+\frac{1}{\hat{Q}} + \frac{p}{\sqrt{\hat{Q}}}+\frac{1}{\sqrt{\hat{Q}M}}\right) \vspace{-1.5mm} $$}
\noindent we observe the experimental results in Figure~\ref{fig:fix_rep} and \ref{fig:hyper} of Sections~\ref{sec:FL} and~\ref{sec:performance} respectively, when varying the parameters of $N$, $p$, $a$ and $\kappa$, results are consistent with this error rate.
\end{remark}

\begin{remark}
Derived from the Corollary~\ref{co:corollary1} and Remark~\ref{re:remark1}, there is a trade-off problem between convergence speed and error rate according to the level of reward $\kappa$ and punishment $\eta$ from the reputation model. This trade-off problem is mainly based on the fact that if the model penalises the bad behaviours of clients heavily, it would decrease their reputation dramatically so the model would take a longer time to converge. On the other hand, mitigating the punishment to increase the reward, would lead to an increase in the error rate.
\end{remark}
%\begin{remark}
%(Fixed Malicious clients) The conclusion also demonstrate the significance of assuming the percentage of malicious clients remains constant over rounds. If each round enables a varied proportion of clients to be malicious, then bound of Theorem~\ref{th:theorem1} is inapplicable to all algorithms and convergence. While it is true that attackers may impersonate others or, more broadly execute Sybil attacks, simple procedures such as preregistration of all participants (possibly using some kind of identification) might help to prevent such assaults. 
%\end{remark}
%$$ \frac{1}{T}\sum_{t=1}^{T}\mathbb{E}\left \| \nabla{\mathcal{L}(D,w^{t-1})} \right \|\leq $$

\section{Performance Evaluation}\label{sec:performance}
The objectives of our experimental evaluation are the following: (a) evaluate the performance of our aggregation method against other state-of-art robust aggregation methods, (b) benchmark it in three different scenarios, namely, no attack, label flipping attack, and backdoor attack, (c) do so using a text based real-world dataset of sensitive categories from~\cite{matic2020identifying} to which we will henceforth refer to as SURL, and finally (d) show that our experimental result are consistent with our previous theoretical analysis. 

\subsection{Experimental Setup}
\subsubsection{Datasets} \label{subsub:datasets}
The SURL dataset comes from a crowdsourcing taxonomy in the Curlie project~\cite{Curlie}, containing six categories of URLs: five sensitive categories (Health, Politics, Religion, Sexual Orientation, Ethnicity) and one for non-sensitive URLs, with a total of 442,190 webpages. The number of URLs in sensitive and non-sensitive categories are equally balanced. Each sample contains content, metadata and a class label of the webpage. For the SURL text classification task, we train a neural network with three fully connected layers and a final softmax output layer, same as in the evaluated methods~\cite{fu2019attack,yin2018byzantine}.
%\ty{Although we mainly focus on classifying webpages with sensitive content, for completeness we also evaluate the performance of the proposed FL methods in a different context related to an image classification task using CIFAR-10~\cite{krizhevsky2009learning}. We observe similar results to those reported for the SURL dataset (see Appendix~\ref{app:CIFAR-10}).}
Furthermore, in order to fulfil the fundamental setting of an heterogeneous and unbalanced dataset for FL, we sample $u_k$ from a Dirichlet distribution~\cite{minka2000estimating} with the concentration parameter $\iota = 0.9$ as in~\cite{bagdasaryan2020backdoor}, which controls the imbalance level of the dataset, then assigns a $u_{k, i}$ fraction of samples in class $k$ to client $i$, with the intention of generating non-IID and unbalanced data partitions.
As a sanity check, we also tested our reputation scheme on a different classification task involving images and got consistent results as those we got for sensitive content (see Appendix~\ref{app:CIFAR-10}). 

%% A sample notebook with the results of running our model will be published in a public repository~\footnote{anonymous} with visible and readable figures that reproduce our results.
%\agr{We save the results in the notebook of Colab Pro in a file and publish them. No need to buy Colab Pro, but if we do, the code will have be in google drive to be able to run so i would just publish a checkpoint we save in Colab for now. Reviewers would not look at code but we show we did not make up numbers. For the repo, we have time until January to clean it}.
%\ty{great,then I would update when I have all the results.now the one you shown is missing a lot new results.}

\subsubsection{Threat Model}\label{sec:threat model} 
We consider the following threat model.
\\
\noindent\textbf{Attack capability:}
In the FL setting, the malicious clients have complete control over their local training data, training process and training hyper-parameters, e.g., the learning rate, iterations and batch size. They can pollute the training data as well as the parameters of the trained model before submitting it to the server but cannot impact the training process of other clients. We follow the common practice in the computer security field of overrating the attacker’s capability rather than underrating it, so we limit our analysis to worst-case scenarios. There, an attacker has perfect knowledge about the learning algorithm, the loss function, the training data and is able to inspect the global model parameters. However, attackers would still have to train with the model published by the server, thus complying with the prescribed training scheme by FL to their local data. Furthermore, the percentage of byzantine clients $p$ is an important factor that determines the level of success for the attack. We assume that the number of attackers is less than the number of honest clients, which is a common setting in similar methods~\cite{yin2018byzantine, fu2019attack} to the ones we evaluate and compare our method with.
\\
\noindent\textbf{Attack strategy:} We focus on two common attack strategies for sensitive context classification, namely, (i) label flipping attack~\cite{barreno2010security} and (ii) backdoor attack~\cite{bagdasaryan2020backdoor}. Comparing to other attacks, for example model poisoning attack~\cite{shejwalkar2021manipulating,fang2020local}, 
these two data poisoning attacks are more likely to be carried out by real users in the real world via our browser extension described in Section~\ref{sec:EITR}, since polluting data is easier than manipulating model updates using the browser extension. Note that privacy attacks including membership inference attack~\cite{naseri2020local} and property inference attack~\cite{melis2019exploiting}, are out of the scope of this paper, but form part of our ongoing and future work.

In a label flipping attack, the attacker flips the labels of training samples to a targeted label and trains the model accordingly. In our case, the attacker changes the label of ``Health'' to ``Non-sensitive''. In a backdoor attack, attackers inject a designed pattern into their local data and train these manipulated data with clean data, in order to develop a local model that learns to recognise such pattern. We realise backdoor attacks inserting the top 10 frequent words with their frequencies for the ``Health'' category.
%and a special pixel pattern~\cite{fu2019attack} to CIFAR-10 dataset at training time. 
Therein the backdoor targets are the labels ``non-sensitive''. 
%and ``bird'' respectively
A successful backdoor attack would acquire a global model that predicts the backdoor target label for data along with specific patterns. 

For both attacks, instead of a single-shot attack where an attacker only attacks in one round during the training, we enhance the attacker by a repeated attack schedule in which an attacker submits the malicious updates in every round of the training process. Also, we evaluate a looping attack where the attackers spreads out poisoning every 30 epochs for the label flipping attack based on Figure~\ref{fig:rep_loop}. It is important to note that even if the attackers have full knowledge of our method, they would still be unable to mount smarter attacks that would try to maximise the damage caused while minimising their reputation drop. This is because the attackers are unaware and cannot compute their reputation score since the latter is computed at the server and requires input from all clients. Moreover, we allow extra training epochs for an attacker, namely, being able to train the local models with 5 more epochs as in~\cite{bagdasaryan2020backdoor}.

\subsubsection{Evaluated Aggregation Methods}
\label{subsec:evaluatedmethods}
We compare the performance of our aggregation method against the existing state-of-the-art in the area FedAvg~\cite{mcmahan2017communication}, as well as against popular robust aggregation methods such as Coordinate-wise median~\cite{yin2018byzantine}, Trimmed-mean~\cite{yin2018byzantine}, FoolsGold~\cite{fung2018mitigating,fung2020limitations}, Residual-based re-weighting~\cite{fu2019attack}, and FLTrust~\cite{cao2021fltrust}.

\noindent\textbf{FedAvg} is a FL aggregation method that demonstrates impressive empirical performance in non-adversarial settings~\cite{mcmahan2017communication}. Nevertheless, even a single adversarial client could control the global model in FedAvg easily~\cite{blanchard2017machine}. This method averages local model updates of clients as a global model update weighted by the fraction of training samples size of each client compared to total training samples size. We use it as baseline evaluation to assess the performance of our method. \\
\noindent\textbf{Median} is using coordinate-wise median for aggregation. After receiving the updates %$\left \{ \mathbf{w}_{i}^{t} \right \}_{i=1}^{M}$
in round $t$, the global update is set equal to the coordinate-wise median of the updates,
%$ \mathbf{w}^{t} :=  \underset{i}{\mathrm{median}}\left \{ \mathbf{w}_{i}^{t} \right \}$ 
% with its $n$-th coordinate % $\mathbf{w}^{t}_{n} := \underset{i}{\mathrm{median}} \left \{ \mathbf{w}_{i,n}^{t} \right \}$ for each $n \in N$
where the median is the 1-dimensional median. \\
\noindent\textbf{Trimmed-mean} is another coordinate-wise mean aggregation technique that requires prior knowledge of the attacker fraction $\beta$, which should be less than half of the number of  model parameters. For each model parameter, the server eliminates the highest and lowest $\beta$ values from the updates before computing the aggregated mean with remaining values. \\
\noindent\textbf{FoolsGold} presents a strong defence against attacks in FL, based on a similarity metric. Such approach identifies attackers based on the similarity of the client updates and decreases the aggregate weights of participating parties that provide indistinguishable gradient updates frequently while keeping the weights of parties that offer distinct gradient updates. It is an effective defence for sybil attacks but it requires more iterations to converge to an acceptable accuracy.
%\agr{is this correct or the other way around? if true, this would mean FoolsGold penalises similar users. Can explain why they did that or rationale behind? is this in contradiction or contraception with our approach, why?}
%\ty{they assume attackers use similar data so have similar updates, which is different from our case}
%\agr{Yes I know, but why they do it or what is the rationale behind that choice? do we know? If we do, we can justify why we do it differently or not, and why is better or not (threat model) - also missing btw.}
%\ty{As I said the logic behind their choice is they assume attackers use similar data so have similar updates. they didn't consider client's reputation, that is different }.
%\todo{We need to know what is our threat model at least and if it is comparable or not to FoolsGold, and why yes or why not in terms of security.}
\\
\noindent\textbf{Residual-based re-weighting} weights each local model by accumulating the outcome of its residual-based parameter confidence multiplying the standard deviation of parameter based on the robust regression through all the parameters of this local model. In our reputation-based aggregation method, we implement the same re-weighting scheme IRLS~\cite{wilcox2011introduction} as residual-based aggregation, but choose the collection of reputation as the weights of clients' local models. \\
\noindent\textbf{FLTrust} establishes trust in the system by bootstrapping it via the server, instead of depending entirely on updates from clients, like the other methods do. The server obtains an initial server model trained on clean root data. Then, depending on the cosine similarity of the server model and each local model, it assigns a trust score to each client in each iteration.

\subsubsection{Performance Metrics}
We use the average accuracy (Avg-ACC) of the global model to evaluate the result of the aggregation defence for the poisoning attack in which attackers aim to mislead the global model during the testing phase. The accuracy is the percentage of testing examples with the correct predictions by the global model in the whole testing dataset, which is defined as: %\vspace{-1mm} 
$$ \textrm{Avg-ACC} = \frac{\mbox{\# correct predictions}}{\mbox{\# testing samples}}$$ 
In addition, there is existence of targeted attacks that aim to attack a specific label while keeping the accuracy of classification on other labels unaltered. Therefore, instead of Avg-ACC, we choose the attack success rate (ASR) to measure how many of the samples that are attacked, are classified as the target label chosen by a malicious client, namely: %\vspace{-1mm}
$$\textrm{ASR} = \frac{\mbox{\# successfully attacked samples}}{\mbox{\# attacked samples}}$$
A robust federated aggregation method would obtain higher Avg-ACC as well as a lower ASR under poisoning attacks. An ideal aggregation method can achieve 100\% Avg-ACC and has the ASR as low as the fraction of attacked samples from the target label.

\subsubsection{Evaluation Setup}
For the malicious attack, we assume that 30\% of the clients are malicious as in~\cite{blanchard2017machine}, which is also a common byzantine consensus threshold for resistance to failures in a typical distributed system~\cite{castro1999practical}. For the server-side setting, in order to evaluate the reliability of the local model updates sent by the client to the server, we assume that the server has the ability to look into and verify the critical properties of the updates from the clients before aggregating. %\ci{I suggest we remove the following sentence since user privacy techniques other than the FL approach itself is out of scope in this work. We only increase the area of attack here... Probably this sentence trigger the review comments related to users' privacy.} Furthermore, there are some secure techniques to preserve the privacy of client updates by disguising them with noise~\cite{bonawitz2017practical}. In order to employ our detection scheme by tracking fraudulent updates, we assume no secure techniques on the server-side.\ty{Rather than addressing privacy protection, we emphasise the robustness of our method. In future work we will explore aggregating updates with differential privacy or similar privacy methods.} \ci{Again, here we add this additional text to address the review comments related to users' privacy but we use FL here for exactly that reason. Is a contradicting statement in my opinion. We use FL to protect users privacy as we state in the 3rd paragraph in introduction. This is part of our motivation to use FL.}  %\ci{Not clear to me what the above paragraph is trying to explain...} \nla{Same here. I cant understand what the previous paragraph is saying and why?. Looks like something that needs to be removed}

Also, we only consider FL to be executed in a synchronous manner, as most existing FL defences require~\cite{blanchard2017machine,yin2018byzantine,xie2019zeno,fang2020local}. For all the above aggregation methods under attack, 
we perform 100 iterations using the SURL dataset with a batch size of 64 and 10 clients. Furthermore, we evaluate our method for increasing numbers of clients. These settings are inline with existing state-of-the-art methods for security in FL~\cite{bagdasaryan2020backdoor,yin2018byzantine,fu2019attack}
%, unless otherwise stated. 
More details related to the training setting are presented in Appendix~\ref{app:experimental_setting}.
%Later we also explore the impact of varying the percentage of attackers.
 %\cite{yin2018byzantine}\cite{xie2019zeno}\cite{fang2020local}.
%\todo{Right but not sure of how to write this one to make sense, probably with pseudonyms, anonymous ids, local differential privacy. Anyhow, we should integrate that apart. In the Background i say we just assume privacy, which could be enough unless we already know which privacy method would be compatible with our security method.}
% In the future, we will evaluate our technique using differential privacy to supplement our research of Federated learning's security. 

%Due to the assumption that the server synchronizes with clients in every iteration, clients send back their local updates to the server. 

\subsection{Convergence and Accuracy}

\begin{figure*}[!htbp]
     \centering
      \includegraphics[width=\textwidth]{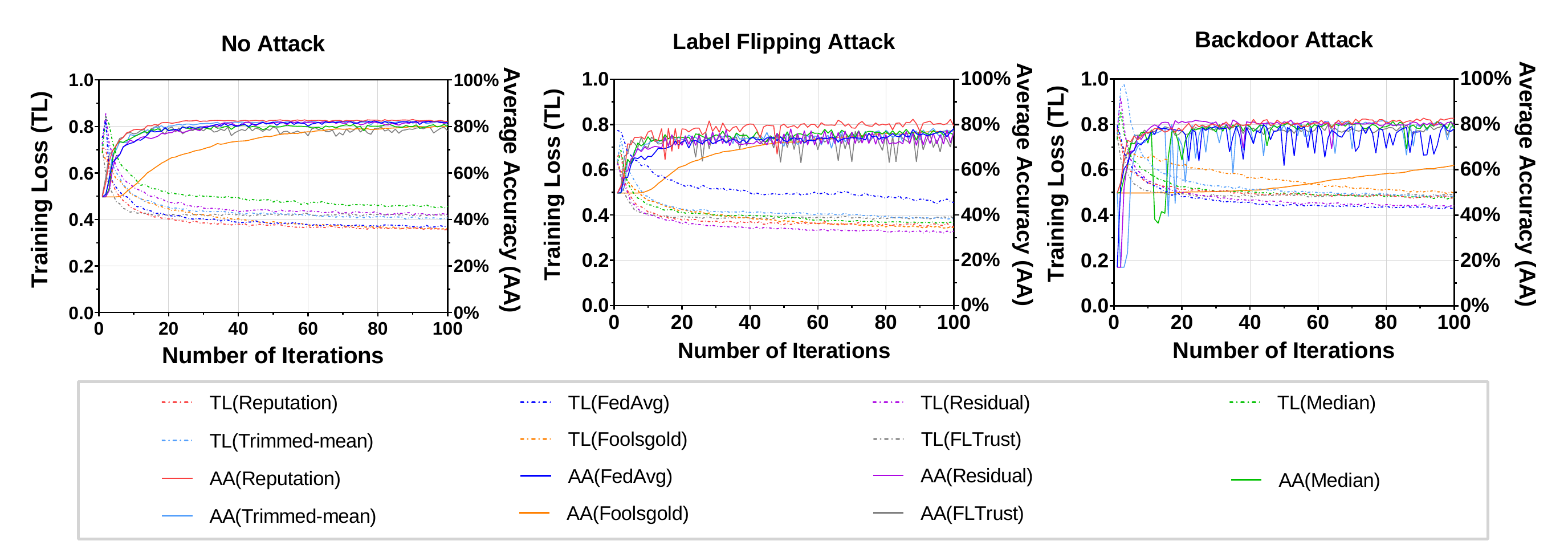}
     \vspace{-6mm}
     \caption{Training Loss (TL) and Average Accuracy (AA) for 100 epochs of Reputation-based, FedAvg, Residual-based, Median, Trimmed-mean, FoolsGold and FLTrust methods in SURL Dataset under no attack (left) scenario, under label flipping attack (middle) and backdoor attack (right) scenarios with 30\% malicious clients.
     }
     \label{fig:SURL_conv}
\vspace{-2em}
\end{figure*} 

In Figure~\ref{fig:SURL_conv} (left), we analyse the performance of our method in the no attack scenario and compare the convergence and accuracy of our method with others during training. We show the training loss (left axis) and average accuracy (right axis) during 100 training iterations for 7 methods.% under the SURL dataset. \ci{Since now we use just one dataset we do not need to name it. We just need to describe it and that's it (We already did in~\ref{subsub:datasets}). We can also remove the dataset name from the figures' caption to avoid confusion.}

Our aggregation starts with the lowest training loss and maintains it throughout the training process. It only takes 24 iterations to achieve 82\% accuracy and then converge to 82.13\%, which represents a 2.7$\times$ faster converge rate than FedAvg. In comparison, Residual-based and Trimmed-mean  have almost identical training loss and take 52 and 47 iterations to reach 82\% accuracy and practically converge to 81.79\% and 80.76\% respectively, which is 2.2$\times$ and 2$\times$ slower than our reputation method.
Median reaches 81\% at 83 rounds and after that converges to 79.94\%, which amounts to a 3.6$\times$ slower converge rate than our method. %\agr{remove next sentence, I can make FoolsGold converge faster in my setup increasing the learning rate. This seems obviously a problem of FoolsGold or how you configure it to run, and yet not clear why ours is faster to converge so reviewers will easily make a guess like my observation. Also, to be consistent with Figures 7 and 8 where we removed FoolsGold, we should not show it here either.} 
Especially, Foolsgold and FLTrust are slow to converge and do not converge within 100 iterations, so our convergence rate is at least 4.2$\times$ better than FoolsGold and FLTrust. %Compared to other approaches, our reputation-based method converges substantially faster and achieves higher accuracy. 
This demonstrates that our reputation model benefits from convergence speed and accuracy performance. This is because our reputation scheme assigns higher weight to more reliable clients when there is no ongoing attack, which generates more consistent updates thereby accelerating the convergence.

\subsection{Resilience to Attacks}
We begin by analysing the performance with a static percentage (30\%) of attackers, and then move on to the performance with a varying percentage of attackers under label flipping and backdoor attacks.
%\nla{I agree with Costas, but to avoid breaking the readability of the text I suggest you put the parameters of the evaluatiopn setting used in the caption of the plot. Since you have already explained the eval setting nicely in the begining you should be able to give the parameters without a lot of text. E.g., Comparisson of X and Y in terms of metric Z. ParamName1=ParamValue1, ParamName2=ParamValue2, ... In the text just keep what you want the reader to remember after reading the section. We compared with X Y Z and we are better} The results show that our method is resistant to these two types of adversarial attacks, with greater accuracy and a lower attack success rate than other evaluated methods.

% Label Flipping
\subsubsection{Label Flipping Attack}
\noindent\textbf{Static percentage of attackers:}
Figure~\ref{fig:SURL_conv}~(middle) shows the convergence of mentioned methods under label flipping attack. Our method converges 1.8$\times$ to  2$\times$ faster than all competing state-of-the-art methods under  attack, enlarging its performance benefits compared to the no attack scenario.
In addition, our method outperforms competing methods by at least 1.4\% in terms of accuracy.
% \begin{figure}[t]
%     \centering
%     \includegraphics[width=1\columnwidth]{Figure/converge_labelflip.png}
%     % \vspace{-8mm}
%     \caption{Training Loss (TL) and Average Accuracy (AA) of Reputation, FedAvg, Median, FoolsGold and Residual with 30\% malicious clients under label flipping attack over 100 iterations.}
%     \label{fig:converge_labelflip}
% \end{figure}
% When it comes to ASR, the performance of aggregation methods in CIFAR-10 have higher ASR in most circumstances than SURL under label flipping attack. As we utilise the identical label flipping strategy for both datasets, this demonstrate that label flipping attacks are more effective for CIFAR-10 dataset. The reason for this is the difference in model structure between the two datasets. Because of the more complicated model structure employed in the CIFAR-10 dataset, which allows it to learn relatively successfully from collusion data produced by label flipping, the label flipping attack may mislead model effectively and inflict more severe damages to CIFAR-10 than SURL. 
 \begin{figure*}[!bpt]
      \centering
       \includegraphics[width=\textwidth]{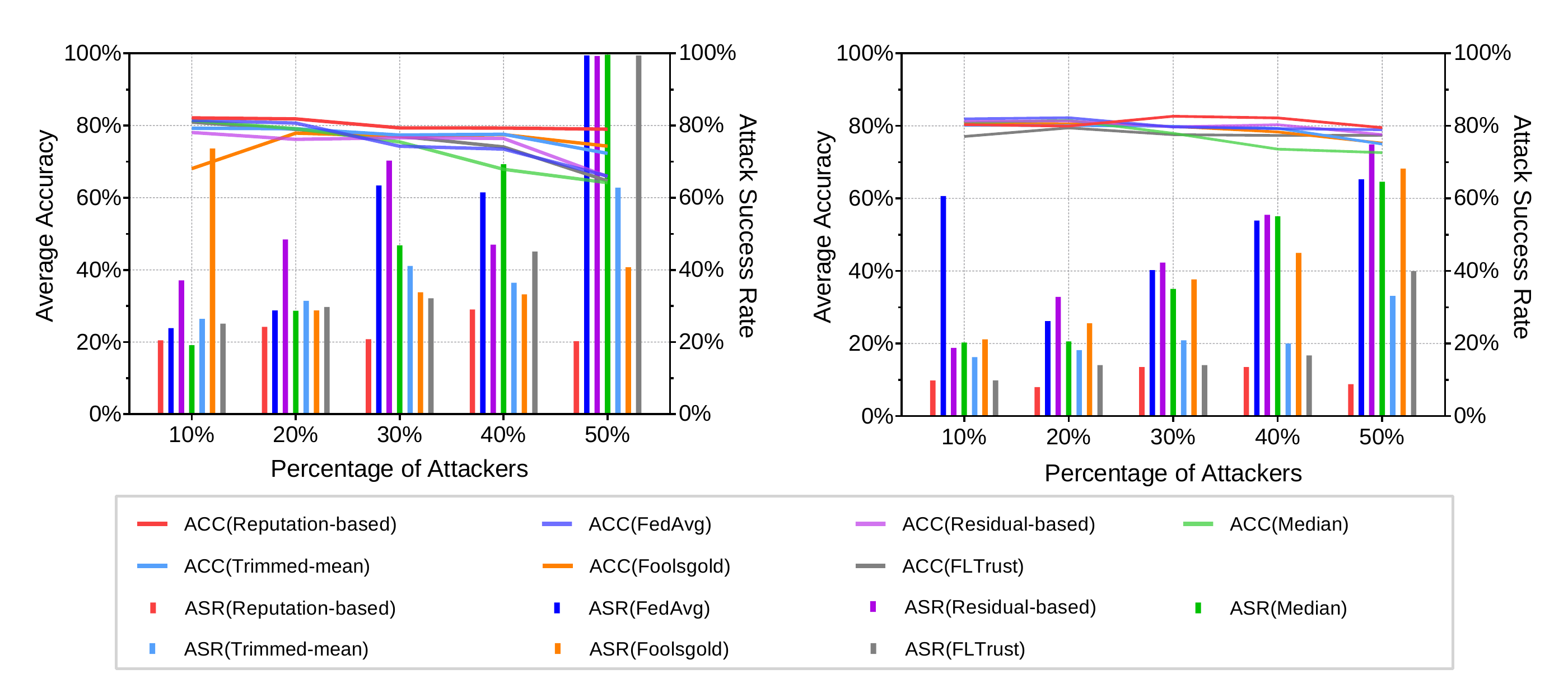}
       %\vspace{-6mm}
      \caption{Average accuracy (ACC) and attack success rate (ASR) for varying percentage of attackers from 10\% to 50\% under label flipping (left) and backdoor (right) attack for Reputation-based, FedAvg, Residual-based, Median, Trimmed-mean, FoolsGold and FLTrust methods in SURL Dataset.}
      \label{fig:SURL_attack}
\vspace{-2em}
 \end{figure*} 
\\
\noindent\textbf{Varying the percentage of attackers:} Here we analyse the impact on our aggregation method as the proportion of attackers increases.
%to observe how it affects the model's performance.
Figure~\ref{fig:SURL_attack}~(left) shows the change of performance metrics for varying percentage of attackers for seven evaluated methods
% \ci{are we using two different datasets here or is just the SURL dataset?}\agr{Only SURL, clearly this was not proofread yet to address that change we agreed beforehand.} 
%for four evaluated methods except Foolsgold under label flipping attack. The reason that Foolsgold is not included in Figure~\ref{fig:SURL_attack}~(left) is that it performs consistently with 56.53\%-57.65\% Avg-ACC and less than 1\% ASR  % \ci{if we use only the SURL dataset we can remove ``in SURL'' from here.} in SURL, \ci{what are the following numbers refer to? Currently it is not clear. Is the result for the 50\% of attacks?} 9.90\%-10.27\% Avg-ACC, 
%and less than 1\% in CIFAR-10 respectively
%when we increase $p$ ranging from 10\% to 50\% under label flipping attack. It demonstrates that Foolsgold is a strong defence, but needs many more iterations to converge to acceptable accuracy.  %Naturally, as the number of malicious clients increase, the total attack success ratio increases but the accuracy declines. This is because attacks are designed in such a way that the fewer malicious clients there are, the less impact attacks have on the model. As a result, the model is less likely to be attacked, and malicious participants have a reduced success rate in attacks when most of participants are honest. 
When the percentage of attackers $p$ ranges from 10\% to 50\%, our method is resistant against label flipping attacks with a small loss in accuracy and a consistent attack success rate of all the methods.
As $p$ approaches 50\%, FedAvg, Residual-based, Median and FLTrust defences become ineffective in mitigating the attack, and correspondingly their Avg-ACC decreases linearly. Moreover, under label flipping attack during the whole process, our reputation-based method has the highest accuracy outperforming other methods by 1\% to 23.1\%. At the same time it has the lowest ASR. The average ASR of other methods are at least 82.8\% higher than ours.
% Figure~\ref{fig:label_flipping}(b) shows that accuracy only decreases slightly with the exception of the Median approach, and slowly increases the attack success rate during the process of reducing the number of honest clients in CIFAR-10.
% Meanwhile, our reputation-based aggregation has the second lowest attack success ratio, which is quite significant less than others, excludes Foolsgold (0\%).
%Although the SURL model has a basic architecture that causes it to learn from data inefficiently, when the number of attackers reaches a certain threshold, the model will be dominated by collusion data and collapse instantly. 

%In Additionally, although the ASR increases, the Avg-ACC of our method remains stable, slightly decreases, and achieves the highest accuracy among all the methods during the process. 
%In summary, the results under label flipping attack in both scenarios demonstrate that our reputation-based model outperforms all other approaches on the average accuracy  as well as having the second lowest attack success in both datasets.

\subsubsection{Backdoor Attack}
\noindent\textbf{Static percentage of attackers:}
Figure~\ref{fig:SURL_conv}~(right) shows the convergence of mentioned methods under backdoor attack. Same as in no attack and label flipping attack scenario, our method converges 1.6$\times$ to 2.4$\times$ faster than all competing state-of-the-art methods. In addition, our method outperforms competing methods by 3.5\% to 33.6\% in terms of classification accuracy.
\\
\noindent\textbf{Varied percentage of attackers:} We examine the scenario in which the percentage of attackers increases. Figure~\ref{fig:SURL_attack}~(right) shows the performance  for the seven evaluated methods under backdoor attack when varying the percentage of attackers $p$ from 10\% to 50\%. Figure~\ref{fig:SURL_attack}~(right) demonstrates that under backdoor attack, our reputation-based method has a consistent accuracy throughout the process with the lowest attack success rate, whereas the average ASR of other methods is at least 72.3\% higher than ours. As $p$ changes, the ASR of the Residual-based, Median, and Foolsgold methods increase linearly. Although FLTrust has a stable ASR, it increases by a factor of 1.39 when $p$ reaches 50\%.

First, we evaluate a varying compromise rate for the label flipping and backdoor attacks using our reputation-based method. For the label-flipping attack, we vary the percentage of the flipped label poisoned by attackers from 10\% to 90\%. Also, for the backdoor attack,  we vary the number of top frequent words inserted as the trigger pattern, from 5 to 25. The remaining settings are the same as in previous experiments.
Figure~\ref{fig:attack_percent} plots the ACC and ASR when varying the compromise rate for both attacks. Figure~\ref{fig:attack_percent} (left) shows that when the percentage of the poisoned sample is increased, it leads to the decrease of the accuracy of the model and to a slight increase of ASR. Figure~\ref{fig:attack_percent} (right) shows that when we increase the number of frequent words from 5 to 20, the ASR remains unaffected. When the frequent words exceed 25, the attack becomes less stealthy and thus can be more easily detected resulting in a lower ASR.

We also evaluate the performance of our method in terms of the number of participating clients. With a 30\% compromise rate, we expand the number of clients from 10 to 200. Our method performs consistently for a larger number of clients, as seen by the stable ACC and ASR as the number of clients grows in Figure~\ref{fig:num_clients}.

\subsubsection{Analysis of Attacks}
\begin{figure}[bpt]
    \centering
    \includegraphics[width=\columnwidth]{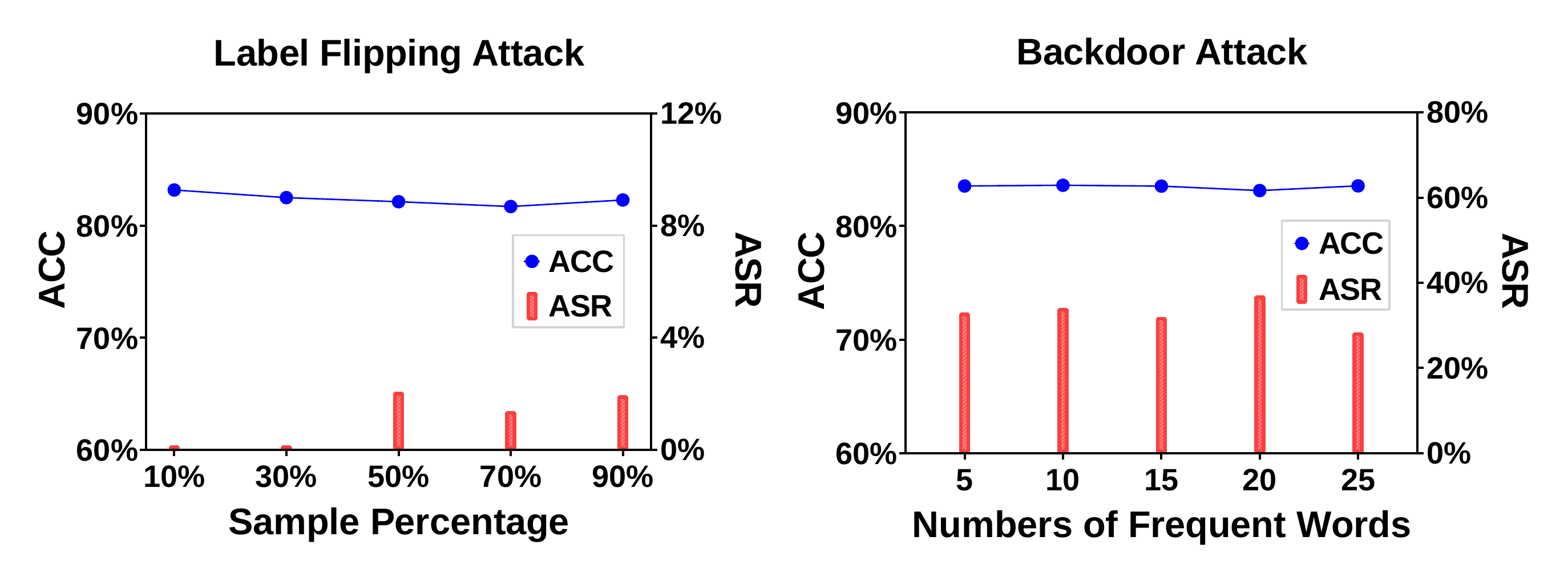}
    %\vspace{-6mm}
    \caption{ACC and ASR as we vary the percentage of flipped label from 10\% to 90\% (left) for label flipping attack, and the number of the frequent words as the trigger pattern from 5 to 25 for backdoor attack(right).}
    \label{fig:attack_percent}
\vspace{-2em}
\end{figure}

\begin{figure}[bpt]
    \centering
    \includegraphics[width=\columnwidth]{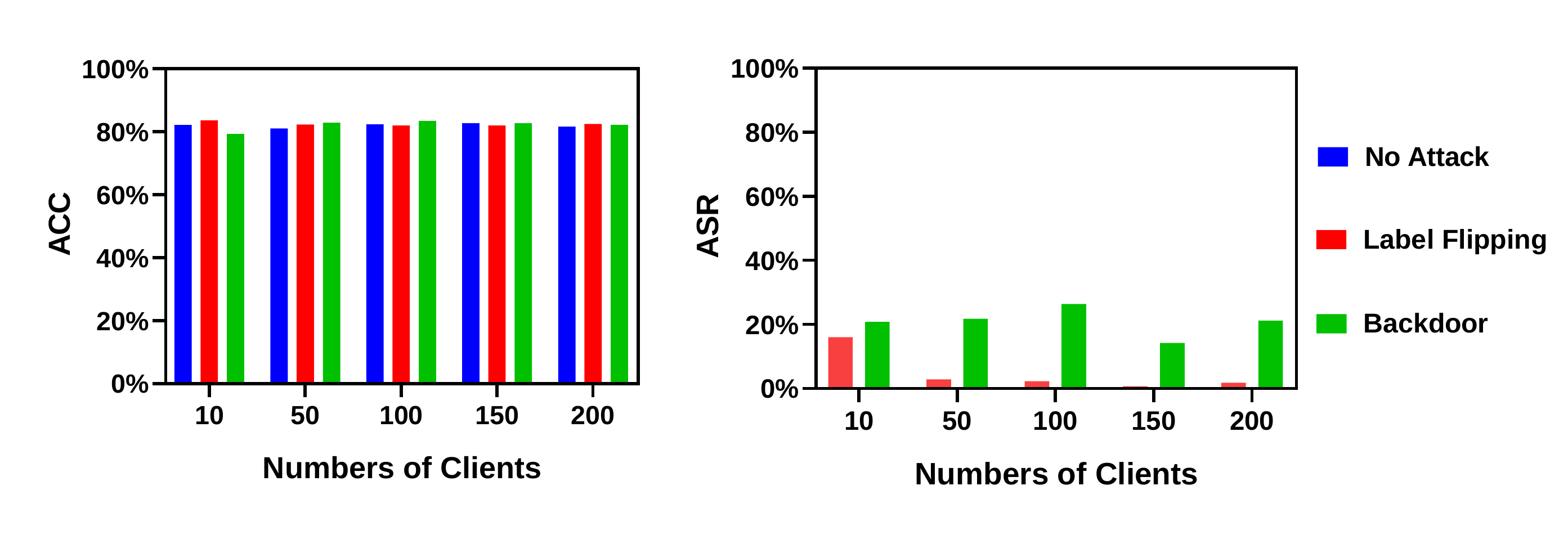}
    %\vspace{-6mm}
    \caption{Average accuracy (ACC) and attack success rate (ASR) for varying the number of clients from 10 to 200 under label flipping  and backdoor attack with 30\% malicious clients.}
    \label{fig:num_clients}
\vspace{-1em}
\end{figure}

Finally, instead of repeating the attack at every epoch, the attacker stretches poisoning across 30 epochs in our study of the looping attack.
The performance of the looping attack is seen in Figure~\ref{fig:loop_attack}. As expected, the looping attack is not as effective as the repeated attack that we previously assessed. All the methods manage to defend it with low ASR, and our method still has the greatest accuracy.

\subsubsection{Evaluation Results}
%\agr{Separate this summary as a subsection, as here we are talking of both Backdoor and Label Flipping but current subsection topic is confusing when reading under Backdoor only. In fact, summarise results of Section 4 all together.} \ci{We can delete the following to save space if needed since we just repeat the main findings of the above subsections. Note that by adding more subsections we waste more space.}
In the no attack scenario, we observe (i) Our method converges 2$\times$ to 4.2$\times$ faster than all competing state-of-the-art methods. (ii) Our method is at least as good or outperforms competing methods in terms of classification accuracy. The above validates that our reputation scheme is helpful even in the no attack scenario. 
This is due to the fact that in our algorithm we give higher weights to the clients with high-quality updates, as illustrated in Figure~\ref{fig:comparision}, causing the model to converge rapidly and retain consistent accuracy. In addition, even under the two different attacks, our method:
%\vspace{-2mm}
\begin{itemize}
[
    \setlength{\IEEElabelindent}{\dimexpr-\labelwidth-\labelsep}% Wrapping of text beyond first line of \item
    \setlength{\itemindent}{\dimexpr\labelwidth+\labelsep}% identation for each new \item
    \setlength{\listparindent}{\parindent}% Restore regular paragraph indentation
]
\setlength{\itemsep}{0pt}
     \item converges  1.6$\times$ to 2.4$\times$ faster than all competing state-of-the-art methods.
    \item provides the same or better accuracy than competing methods. \item yields the lowest ASR compared to all other methods, with the average ASR of them being at least 72.3\% higher than ours.
\end{itemize}
%In conclusion, even under the two different attacks, our method (a) converges between 3.3$\times$ to at least 5.3$\times$ faster than all competing state-of-the-art methods. (b) Outperforms in at least 1\% the competing methods in terms of classification accuracy. (c) Yields an average 3.8\%-15.2\% improvement in ASR against all other methods, with the exception of FoolsGold which is the slowest one to converge and with the lowest accuracy.
%\vspace{-1mm} 
%When compared to other aggregation approaches, such as the Median method, it has relatively low accuracy even when no attackers are present since they lose the majority of the information during model aggregation. In contrast, our method utilises the reweighed average of all local models and therefore gets more information in an unconstrained manner. 
% When comparing our algorithm with the state-of-the-art algorithm Foolsgold, we observe that Foolsgold holds the lowest ASR for SURL and CIFAR-10, as well as the lowest Avg-ACC in both scenarios. The main reason for Foolsgold's performance is due to its inability to distinguish between malicious and clean upgrades. It allocates lower aggregated weights to high similarity of updates, making it resistant to attacks, but at the price of precision.
We obtained comparable findings for the evaluation of the aforementioned methods on 100 clients, as presented in Appendix D.
Furthermore, the result is consistent with the theoretical analysis: as $p$ increases, so does the error rate. % and it also illustrates that $N$ impacts the performance of our aggregation. 
%It is worth mentioning that we lower attack success rate by an average of 20.64\% for backdoor attack and increase accuracy by an average of 3.22\% for backdoor attack when compared to the residual-based technique, which we employ to establish our reputation methods with their residual confident score.

\begin{figure}[!bpt]
    \centering
    \includegraphics[width=\columnwidth]{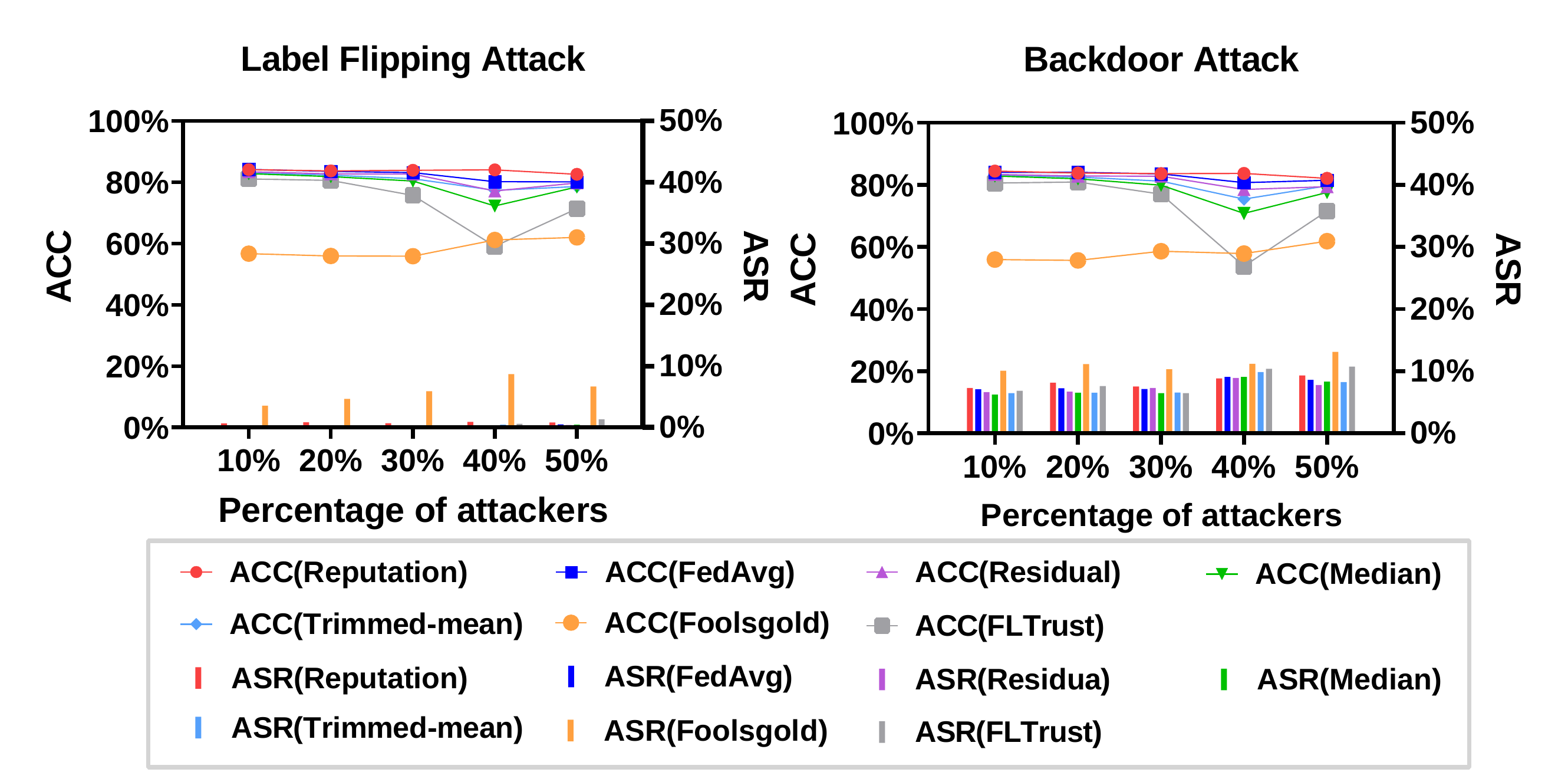}
    %\vspace{-6mm}
    \caption{Average accuracy (ACC) and attack success rate (ASR) for varying percentage of attackers from 10\% to 50\% under label flipping (left) and backdoor (right) attack with a looping attack in which an attacker attacks every 30 epochs.}
    \label{fig:loop_attack}
\vspace{-1em}
\end{figure}

\begin{figure}[bpt]
     \centering
     \begin{subfigure}[b]{0.43\columnwidth}
         \centering
         \includegraphics[width=\columnwidth]{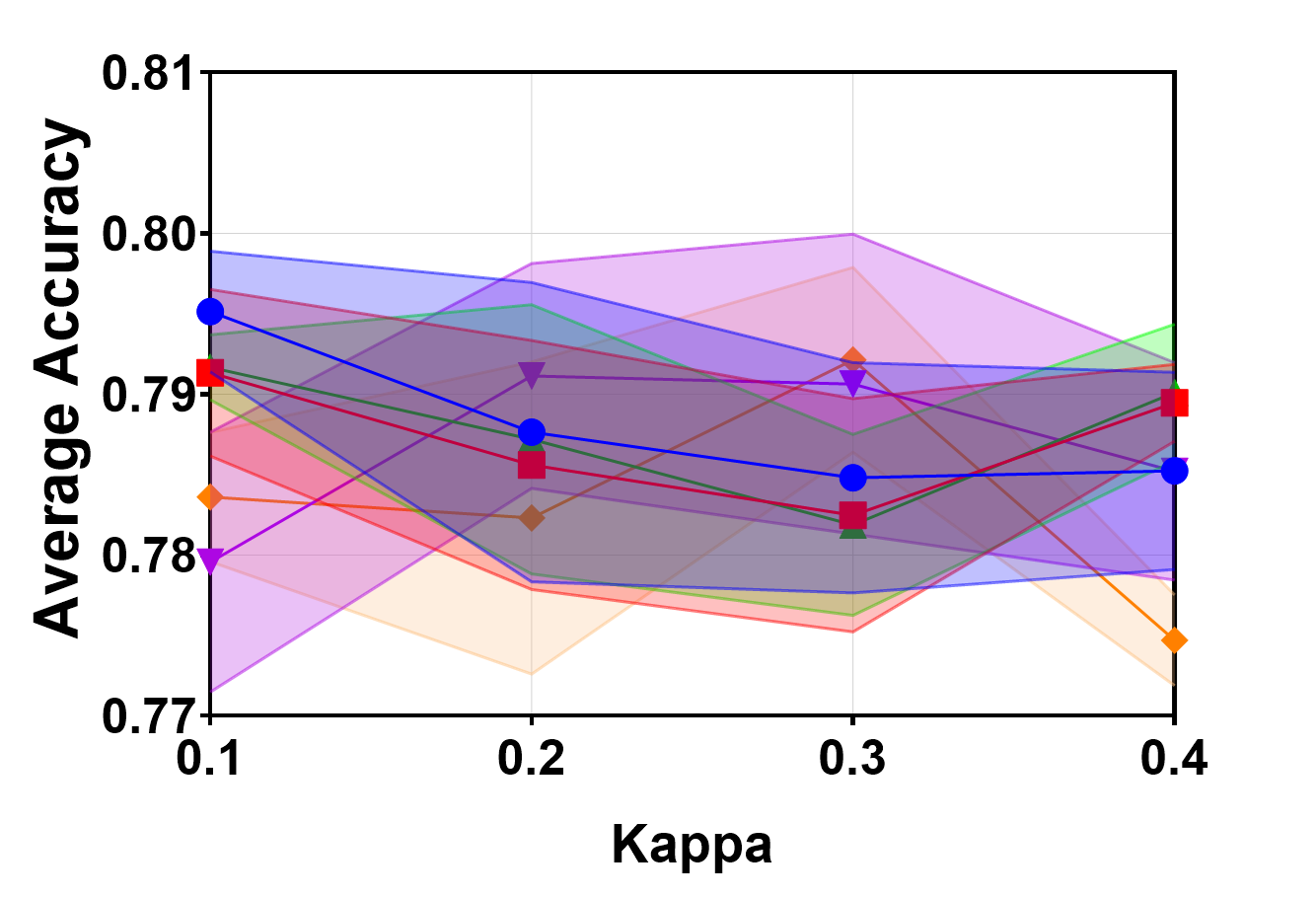}
         \caption{Accuracy}
         \label{fig:Hyper_ACC}
     \end{subfigure}
     \hfill
     \begin{subfigure}[b]{0.5\columnwidth}
         \centering
         \includegraphics[width=\columnwidth]{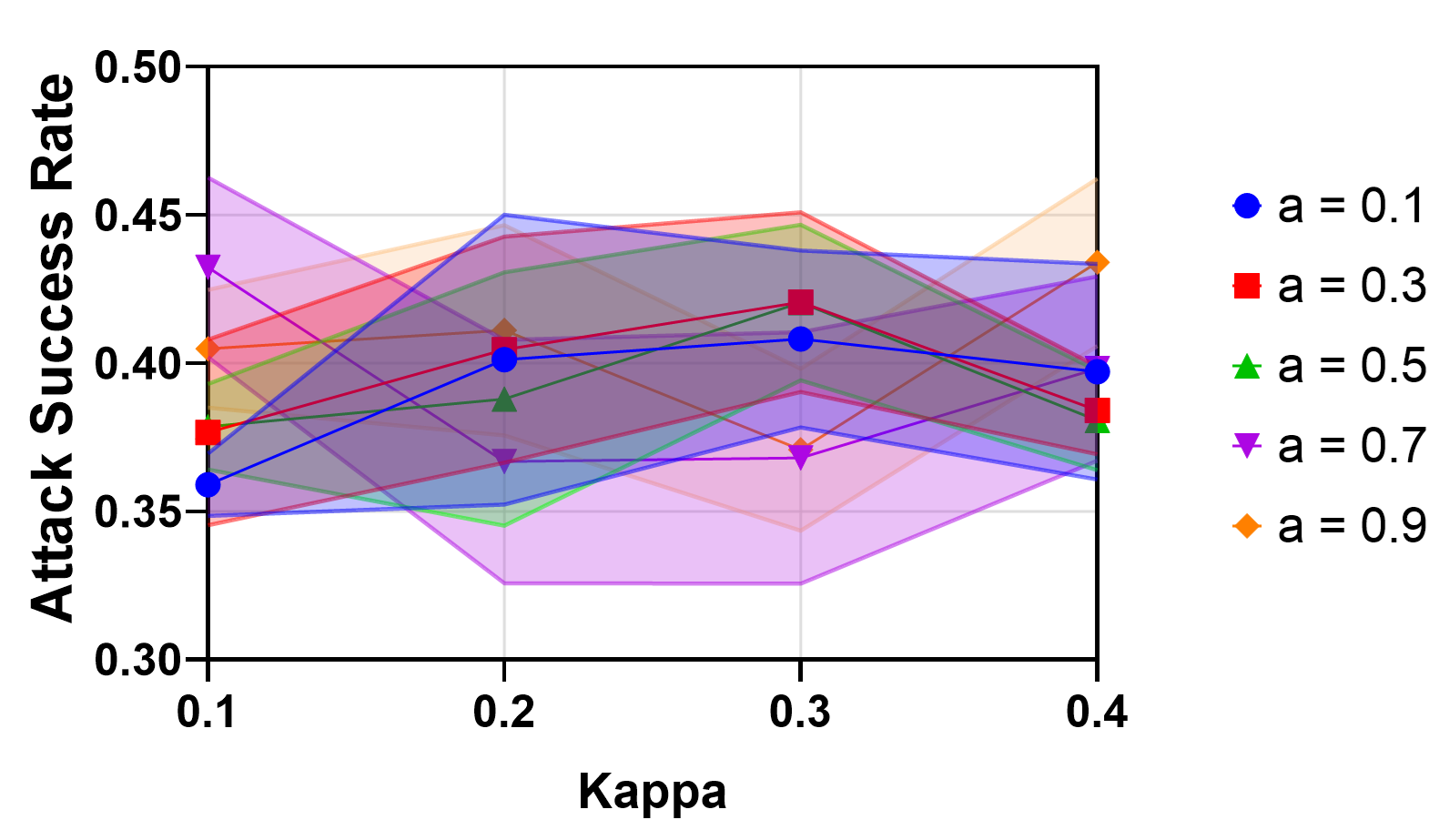}
         \caption{Attack Success Rate}
         \label{fig:Hyper_ARS}
     \end{subfigure}
     \vspace{-1em}
     \caption{Average accuracy and attack success rate as we vary rewarding weight $\kappa$ and prior probability $a$.}
     \label{fig:hyper}
\vspace{-1em}
\end{figure}

\subsection{Stability of Hyper-parameters}\label{app:param-stability}
%In order to demonstrate the impact of model hyper-parameters, we grid search the associated hyper-parameters. 
We employ four hyper-parameters in our reputation model: rewarding weight $\kappa$, prior probability $a$, time decay parameter $c$ and window length $s$.  As shown in Remark~\ref{re:remark1}, $c$ and $s$ do not affect the performance of our model, we only consider hyper-parameters $\kappa$ and $a$, where $\kappa$ controls the reward weight to positive observations and $a$ controls the fraction of uncertainty converted to belief. To demonstrate the impact of these two hyper-parameters of our reputation model, we grid search $\kappa$ in $\left [0.1,0.2,0.3,0.4 \right ]$ and $a$ in $\left [0.1,0.3,0.5,0.7,0.9 \right ]$. The setup is the same as on SURL dataset under label flipping attack. The ultimate accuracy of stability of reputation-based aggregation are shown in Figure~\ref{fig:hyper}. Note that these results are tested for the label flipping attack and they hold according to theory also for backdoor.

The result in Figure~\ref{fig:hyper} demonstrates that our approach is very stable and efficient in terms of hyper-parameter selection, and it achieves a high degree of precision. Furthermore, the result is compatible with the theoretical analysis in Section~\ref{sec:theory}.

\subsection{Comparison against a residual-based method}
To demonstrate how our method improves the residual-base method by assigning the aggregation weights based on reputation, we consider a scenario with 10 clients in the FL system, 8 of which are malicious.
The training lasts 10 communication rounds during which attackers carry out the backdoor attack.
The remaining settings are the same as the default.
Results are shown in Figure~\ref{fig:comparision}, in which the first two clients are benign, and the rest are malicious.
We observe that for our reputation method the aggregation weights of malicious clients, which are their reputations, are rectified to 0 since the second round, demonstrating that our method is successful in eradicating their influence. On the other hand, the aggregate weights of malicious clients in residual-based methods, which are calculated by multiplying the parameter confidence by its standard deviation, are nearly similar and non-zero. This is because repeated median regression seldom yields 0 for the parameter confidence, which causes practically non-zero weights to be assigned to malicious clients by residual-based methods. To address this issue, the reputation model uses positive and negative observations that introduce rewards and punishments to assign divergent weights  to clients. As a result, benign clients are given higher weights whereas malicious clients are eliminated from the aggregation.

\begin{figure}[!bpt]
     \centering
      \includegraphics[width=\columnwidth]{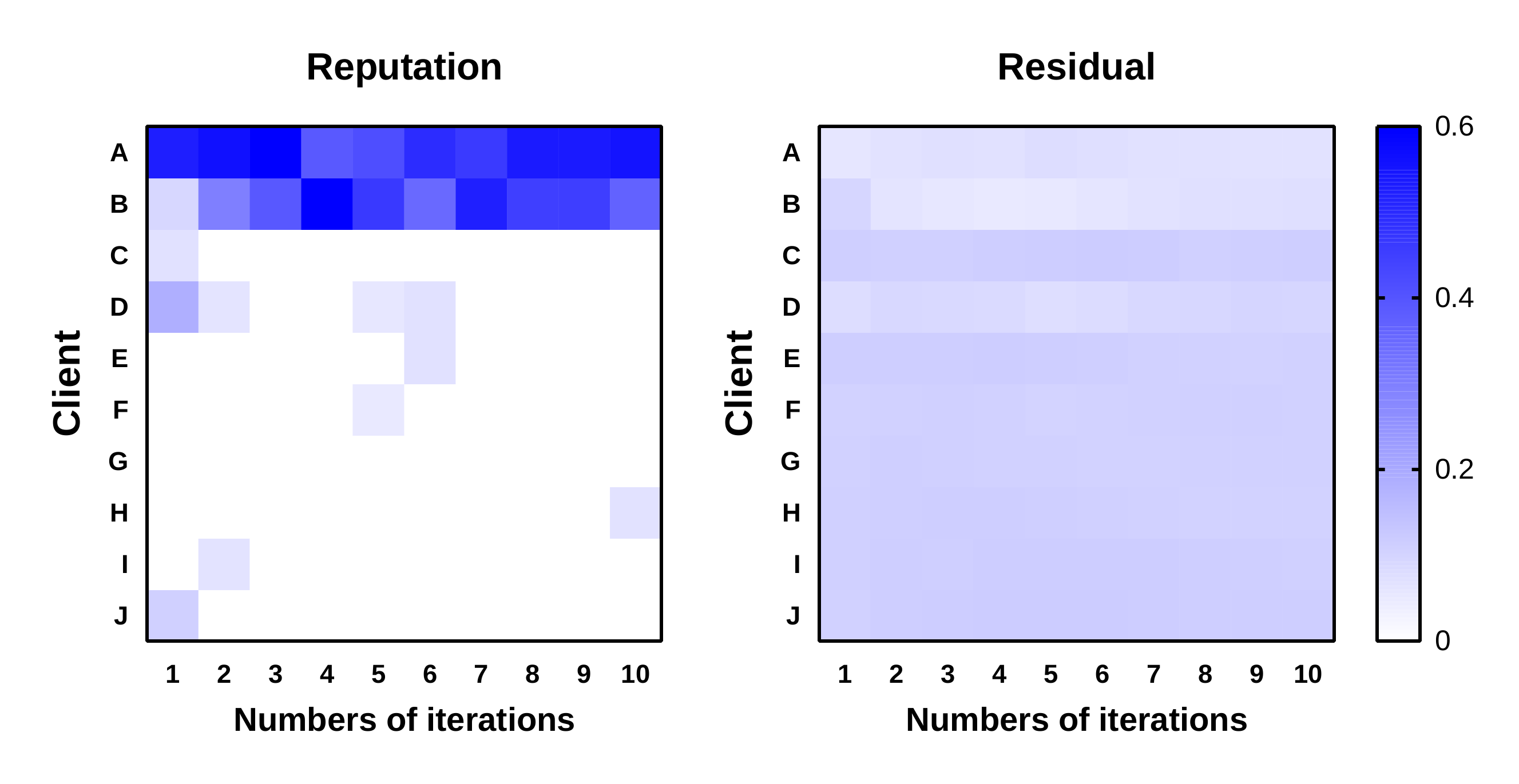}
      \vspace{-1em}
     \caption{The aggregation weights of clients from our reputation-based (left) and residual-based method (right) for 10 communication rounds under label flipping attack with 80\% attackers.
     }
     \label{fig:comparision}
\vspace{-1em}
\end{figure}

\section{The \emph{EITR} system}\label{sec:EITR}
In this section, we provide a high level description of our \emph{EITR}~\cite{eitr} system (standing for ``Elephant In the Room'' of privacy). We then present some preliminary results with real users demonstrating the ability of the system to quickly learn how to classify yet unseen sensitive content, in our case COVID-19 URLs pertaining to the category Health, even in view of inaccurate user input. The system is currently being used as a research prototype to evaluate the robustness of our algorithm in a simple real-world setting. A full in depth description of the system and its performance with more users and more intricate settings, including adoption, incentives, and HCI issues, over a longer time period is the topic of our ongoing efforts and will be covered by our future work.

\subsection{System Architecture and Implementation}

\begin{figure}[!bpt] %[!htbp]
    \centering
    \includegraphics[width=0.8\columnwidth]{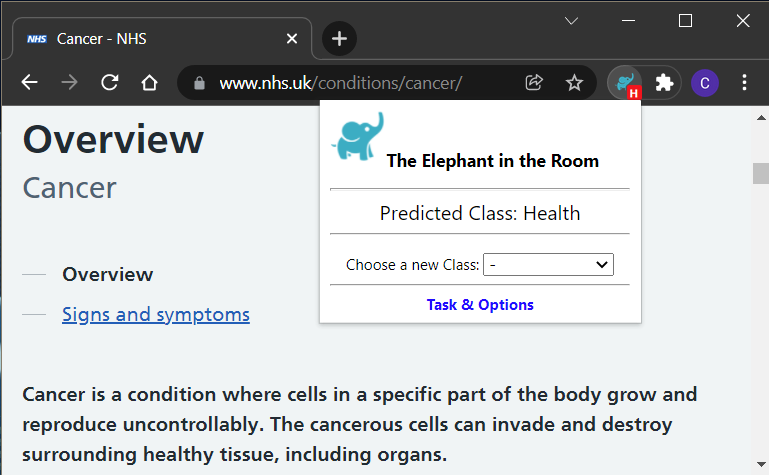}
    \caption{%The Elephant In The Room (\emph{EITR}) \ci{EITR is already defined in introduction.}
    \emph{EITR} extension in action. The letter ``H'' inside the red frame at the bottom right of the extension's icon indicates that a health-related page has been detected.}
    \label{fig:eitr_in_action}
\vspace{-2em}
\end{figure}

The \emph{EITR} system is based on the client-server model. The back-end server is responsible to distribute the initial classification model and the consequent updated model(s) to the clients and receive new annotations from the different clients of the system. The client is in the form of a web browser extension that is responsible to fetch and load the most recent global classification model to the users' browser from the back-end server. The loaded model can then be used to label website in real time into the 5 different sensitive topics as defined by GDPR, i.e., Religion, Health, Politics, Ethnicity and Sexual Orientation. Next, we provide more details for each part of the system.\\
\noindent\textbf{Back-end server:} The back-end server is written in JavaScript using the node.js Express~\cite{express} framework. To build the initial classification model we use the dataset provided by Matic et al.~\cite{matic2020identifying}, and the TensorFlow~\cite{TensorFlow} and Keras~\cite{Keras} machine learning libraries. The final trained model is then exported using the TensorFlow.js~\cite{TensorFlowJS} library in order to be able to distribute it to the system's clients (browser extension). The back-end server also includes additional functionalities such as the creation and distribution of users' tasks, i.e., a short list of URLs that the users need to visit and annotate, and an entry point that collects the resulting users annotations during the execution of the task.\\
\noindent\textbf{Web browser extension:} Currently the browser extension only supports the Google Chrome browser and is implemented in JavaScript using the Google Chrome Extension APIs~\cite{chromeAPIs}. To handle the classification model the extension utilises the TensorFlow.js~\cite{TensorFlowJS} library to load, use, and update the model. The main functionality of the extension is to classify the visited website in real time and provide information to the user related to the predicted class as depicted in Figure~\ref{fig:eitr_in_action}. The website classification is based on the metadata (included in the website \texttt{<head>} HTML tag) and the visible text of the website. The extension also allows users to provide their input related to their agreement or disagreement with the predicted class using a drop down list as depicted in Figure~\ref{fig:eitr_in_action} with the label \textit{``Choose a new Class''}.

\subsection{Real Users Experimental Setup}
%Our experimental setup is relatively simple. 
The goal of the real-user experiment is to evaluate our federated reputation-based method on real user activity (instead of systematic tests), and demonstrate that even with real users with different comprehensions of the definition of sensitive information, our method can learn new content fast and achieve higher accuracy than centralised classifiers, which is compatible with our simulation experiment. 

\begin{figure*}[t]
    \centering
    \includegraphics[width=1.0\textwidth]{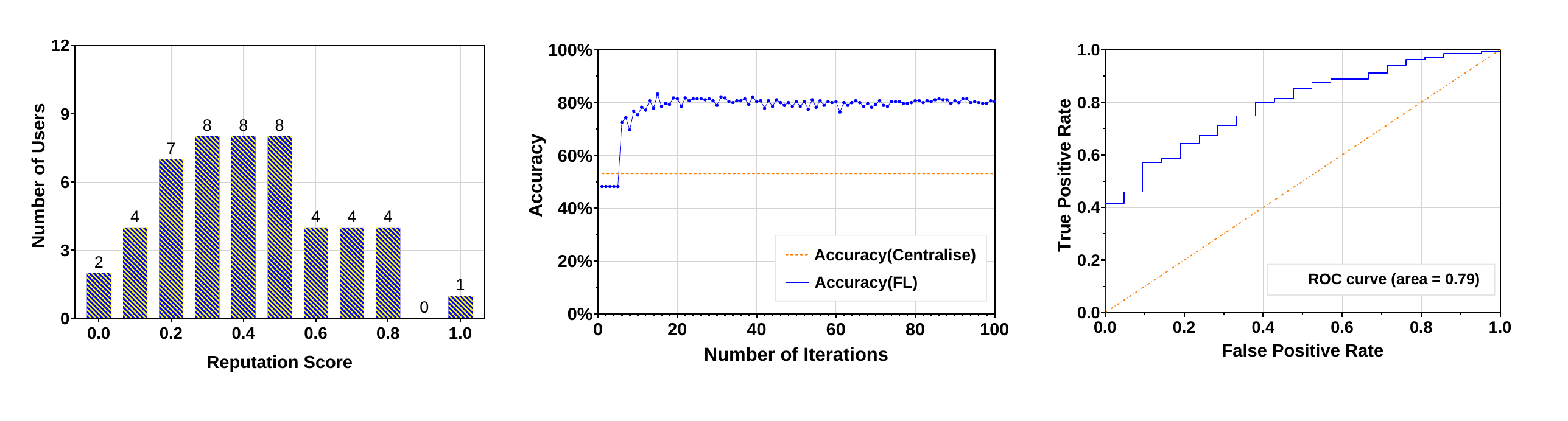}
    %\vspace{-8mm}
    \vspace{-3em}
    \caption{Results of real-user experiment for COVID-19 related URLs with 50 users over 100 iterations. 
    (left): the reputation score of the real users at the end of experiment, 
    (middle): the accuracy of the centralised classifier and the global model of our reputation-based FL approach for COVID-19 URLs, and
    (right): the ROC curve of real-user experiment with 0.79 area under the ROC curve (AUC).}
    \label{fig:real_user}
\vspace{-2em}
\end{figure*}

For the setup, the participant is directed to visit the experiment website that provides the necessary information and instructions on the goal of our study, the definition of sensitive information provided by %\ci{We already define what GDPR is. We can remove the definition here to save space.} 
the current GDPR and how to participate in our study. In order to have access to the browser extension and the installation instructions the user must in advance give explicit consent and accept the data privacy policy. Upon successful installation of the extension, new users are asked to provide a valid email address (to contact them for the reward) and then receive their task, a list of 20 URLs, that they need to visit and provide their labels in order to successfully complete their task. The list of 20 URLs is sampled by the Dirichlet distribution with $\iota=0.9$ for each participant from a database, which includes 300 URLs with sensitive and non-sensitive content related to COVID-19.

\noindent\textbf{Ethical Considerations:}
%\subsection{Ethical Considerations}
We have ensured compliance with the GDPR %\ci{use the GDPR acronym to save space.} 
pertaining to collecting, handling, and storing data generated by real users. To that end, we have acquired all the proper approvals from our institutions. Furthermore, the participants are directed to visit a pre-selected set of URLs selected by us to avoid collecting the actual visiting patterns of our users. In addition, the user input is only collected \textit{if and only if} the user explicitly provide input to the drop-down list labelled \textit{``Choose a new Class''} to avoid collecting the visiting patterns of the user accidentally while they are executing their tasks. Finally, we only use the users' email address to contact them for the reward. The mapping between the user input and their email address is based on a random identifier that is generated during the installation time of the extension.

\subsection{Validation with Real Users}\label{sec:covid-validation}

\noindent\textbf{Data collection:} We  had 50 users participating in our experiment. %These updates are incorporated into the FL model we implement using our aforementioned reputation design. 
In order to evaluate our reputation-based FL method using real-user data, we define a methodology to label ground truth on COVID-19 related sites. 
\\
\noindent\textbf{Ground truth methodology:}
To set the ground truth for COVID-19 sites related to our sensitive or non-sensitive labels, %(e.g, it could be health politics, religion, ethnicity), 
we create a database of 100 websites, which we collect by searching on Google with the query ``sensitive websites about COVID-19'' and choose the top 100 sites returned from the query. Then, four experts in the privacy field, independently annotate them based on their professional expertise in order to achieve an agreement on whether each of those sites included sensitive or non-sensitive content. 
%We let users choose among the categories of Health and non-sensitive for this, allowing users to label a COVID19 related site as Health or Clear.
% Note COVID19 is not a category of the SURL dataset, but a subcategory of its Health category. Given that the topic of COVID19 has been controversial lately, collecting such COVID19 labels can lead to unreliable raters adding updates to the model when using our extension used in a federated manner. 
%The idea here is to show this is something our FL model can also handle, assuming a minimal bootstrapping of reliable groundtruth classification on GDPR-sensitive sites related to newly created Health content, in this case COVID19.
%\ci{The above paragraphs reads a little bit weird. Needs rewriting...}
%In order to do so we provided raters with a simple set of guidelines to follow during annotation as follows:
\\
\noindent\textbf{Ground truth annotation:}
In order to evaluate the annotation of the 100 websites from human experts, we calculate the inter-rater agreement among them using Fleiss Kappa~\cite{fleiss1971measuring}. % The resulting data acts as the ground truth. 
%which returns results of a controversial topic for user privacy as is COVID19~\cite{magge2021overview} due to its Health nature. We collect the first 100 sites returned from the query.
%for which we would expect disagreement on the containing sensitive content, in particular according to the GDPR category of Health employed in the aforementioned dataset SURL we train our FL model with. 
%To ensure the ground truth we produce is reliable and agreement not due to random chance, we ask four raters to annotated those results returned by the query, and 
We obtain 0.56 of Fleiss Kappa, which is an acceptable agreement because the values of Fleiss Kappa. %\footnote{https://www.real-statistics.com/reliability/interrater-reliability/fleiss-kappa/} \ci{move it to the bibliography and cite it instead of using footnote, to save space.}
above 0.5 are regarded as good. Furthermore, given that COVID-19 is a controversial issue, it is difficult for humans to agree on what constitutes sensitive content relating to it. Even though, we still attain a valid ground truth of 85 items belonging to the health sensitive category with agreement ratings of at least 0.5. Note that we also classify the above 100 websites using the centralised classifier proposed in~\cite{matic2020identifying} and get only 53.13\% accuracy.
%We do not consider our pre-trained browser extension as a rater in the human annotation task because we would obtain lower Fleiss Kappa (0.31), which means that we only need to rely in the judgement of our high-quality human experts~\cite{muller2021designing} for  separating chaff from actual COVID19 items into the groundtruth.
\\
\noindent\textbf{Result with real users:} Figure~\ref{fig:real_user} shows the results of accuracy and reputation score with 50 real users in the experiment. Figure~\ref{fig:real_user} (left) shows that the majority of users have reputation scores falling in the intermediate range, with some having a very high reputation and a few having a very low reputation. This indicates the divergence of the user's interpretation of the sensitive information as we expect. In Figure~\ref{fig:real_user} (middle) we compare the accuracy of the centralised classifier and the global model of our reputation-based FL approach for COVID-19 URLs. Despite the diversity of reputation scores of real users, our method converges as rapidly as in simulation and achieves an average accuracy of 80.36\%, thereby verifying the quick convergence and high accuracy results presented in the previous sections. Figure~\ref{fig:real_user} (right) shows that the ROC curve in real-user experiment yielded 0.79 AUC. Our result is acceptable in this scenario because most existing FL techniques are designed to minimise the conventional cost function and are not optimal for optimising more appropriate metrics for imbalanced data, such as AUC~\cite{yuan2021federated}.
%Since our ground truth is made up of about 200 urls related to COVID19 and 100 of Clear category, we add this subset of items to our validation set before training with out newly collected labels from real users in the extension. Therefore, the resulting accuracy takes into account our human-labelled groundtruth for COVID19 related sites.
%\agr{So we will Train with 300 Covid urls of Costas + 100 Clear ones according to Tianyue, and Test with the moreless 100 we have annotated as groundtruth manually by humans that prove to have acceptable inter-rater agreement.}

As we observe, with real users holding our method achieve a good performance. This means that, as new sensitive content appears and/or is defined by GDPR or new upcoming legislation, we will be able to continue training our FL model for this type of task with quick convergence and good accuracy. The empirical results in Figure~\ref{fig:real_user} (middle) also shows that there is a quick convergence to the accuracy's stable value within a small number of iterations (around 30), in line with the theoretical results in Section~\ref{sec:FL}. 

%\subsection{result with real users}
% \begin{figure}
%     \centering
%     \includegraphics[width=0.7\columnwidth]{Figure/ACC_Health.png}
%     \caption{Accuracy of health category in reputation-based FL classifiers and centralised classifiers with 35 users over 100 epochs in real-user experiment}
%     \vspace{5mm}
%     \agr{no need for centralised classifier here but rather the different lines, (i) IMC dataset only for Health and (ii) IMC dataset + new EITR labels for Health. The second (ii) uses our new validation set additionally, while the first (i) did not but can now do use it as well.}
%     \label{fig:ACC_Health}
% \end{figure}

% \begin{figure}
%     \centering
%     \includegraphics[width=0.7\columnwidth]{Figure/reputation.png}
%     \caption{Reputation score of the real users}
%     \label{fig:reputation}
% \end{figure}

\section{Conclusion}\label{sec:conclusions}
In this paper we have shown how to use federated learning to implement a robust to poisoning attacks distributed classifier for sensitive web content. Having demonstrated the benefits of our approach in terms of convergence rate and accuracy against state-of-the-art approaches, we implemented and validated it with real users using our \emph{EITR} browser extension. Collectively, our performance evaluation has showed that our reputation-based approach to thwarting poisoning attacks consistently converges faster than the state-of-the-art while maintaining or improving the classification accuracy. 

We are currently working towards disseminating \emph{EITR} to a larger user-base and using it to classify additional sensitive and non-sensitive types of content. This includes but it is not limited to categories defined by the users themselves for different purposes, not necessarily related to sensitive content, as well as evaluating additional attacks and threat models under our subjective-logic reputation scheme for FL. %and privacy-preserving protections~\cite{zhang2021privsyn} in FL.
In turn, our approach %\ci{replace ``maybe beneficial to other...'' with ``can benefit other...''?} 
can support other FL models going beyond sensitive content classification in future work.

\section*{Acknowledgements}
\noindent We thank the EITR volunteers for their help. This work and dissemination efforts were supported in part by the European Commission under DataBri-X project (101070069), the TV-HGGs project (OPPORTUNITY/0916/ERCCoG/0003) co-funded by the European Regional Development Fund and the Republic of Cyprus through the Research and Innovation Foundation, and the European Research Council (ERC) Starting Grant ResolutioNet (ERC-StG-679158).


% \section*{Acknowledgment}



%\bibliographystyle{IEEEtranS}

%\newpage
\appendix
\subsection{Proofs}
\label{sec:proofs}
The following are the lemmas we utilise in the proof of Theorem~\ref{th:theorem1}.

\begin{lemma}
\label{le:lemma1}
From Assumption \ref{as:assumption1} and \ref{as:assumption4}, $\mathcal L(w)$ is $L$-smooth and $\mu$-strongly convex. Then $\forall w_{1}, w_{2} \in \mathcal{W}$, one has
{\small
\begin{align}
   \langle \nabla\mathcal L(\mathbf{w}_{1})- \nabla\mathcal L(\mathbf{w}_{2})&, \mathbf{w}_{1}  - \mathbf{w}_{2} \rangle  \geq  \frac{L\mu}{L+\mu}\left \|\mathbf{w}_{1}-\mathbf{w}_{2}\right \|_{2}^{2}\notag\\
   & +\frac{1}{L+\mu}\left \| \nabla \mathcal L(\mathbf{w}_{1}) - \nabla \mathcal L(\mathbf{w}_{2})\right \|_{2}^{2}
\end{align}
}
\end{lemma}

\begin{lemma}
\label{le:lemma2}
The difference between $\mathbf{m}(\mathbf{w})$ and $\nabla \mathcal L(\mathbf{w})$ is bounded in every iteration:
{\small
\begin{align}
     \left \|\mathbf{m}(\mathbf{w})-\nabla \mathcal L(\mathbf{w})  \right \|_{2} & \leq  \left \| \mathbf{m}_{0}(\mathbf{w})- \nabla \mathcal L(\mathbf{w})\right \|_{2} + \sqrt{N}\Delta_{1}
\end{align}
}
 
\noindent where: 
$$\Delta_{1} = \frac{M(\varpi(M-1)+\frac{2E}{\sqrt{M}\delta})}{\frac{Wa(M-1)(\kappa N + W)}{(\eta N + W)(\kappa N + Wa)}+1} $$

{\small $$E = \sup \left \{ \frac{37\sqrt{2}\lambda(M+4)}{25(M-1)}
\underset{i}{\mathrm{median}} \left\{ |w_{i,n}^{t}-\hat{B}_{n}x_{i,n}^{t}-\hat{A}_{n}|\right \} \right \} $$}
and $$\mathbf{m}_{0}(\mathbf{w}) := 
\underset{i}{\mathrm{median}} \left \{ \mathbf{m}_{i}(\mathbf{w}) \right \} $$
\end{lemma}
%\proofname 

% \begin{lemma}
% \label{le:lemma2}
% From Assumption \ref{as:assumption1} and \ref{as:assumption4}, $\mathcal L(w)$ is $L$-smooth and $\mu$-strongly convex. Then $\forall w_{1}, w_{2} \in \mathcal{W}$ one has, $$2\mu[\mathcal L(w)- \mathcal L(w^{*})]\leq \left \| \nabla \mathcal L(w) \right \|_{2}^{2}$$ 
% \end{lemma}
% \proofname 

\subsubsection{Proof of Lemma \ref{le:lemma1}}
Let $g(\mathbf{w}) = \mathcal L(\mathbf{w})-\frac{\varsigma}{2}\left \| \mathbf{w} \right \|_{2}^{2}$. Base on the assumption \ref{as:assumption4}, we have $g(\mathbf{w})$ is $(L-\varsigma)$-strongly convex.
from \cite{bubeck2014convex} 3.6, we have
{ \small
\begin{align}
    \langle \nabla\mathcal L(\mathbf{w}_{1})- \nabla\mathcal L(\mathbf{w}_{2}),& \mathbf{w}_{1}  - \mathbf{w}_{2} \rangle  \geq  \frac{1}{L}\left \| \nabla \mathcal L(\mathbf{w}_{1}) - \nabla \mathcal L(\mathbf{w}_{2})\right \|_{2}^{2}
\end{align}
}

\noindent Hence,
{\small
\begin{align}
     \langle \nabla g(\mathbf{w}_{1})- \nabla g(\mathbf{w}_{2}), \mathbf{w}_{1}  - \mathbf{w}_{2} \rangle  \geq \frac{1}{L-\varsigma}\left \| \nabla g(\mathbf{w}_{1}) - \nabla g(\mathbf{w}_{2})\right \|_{2}^{2}
\end{align}
}

\noindent Now We have
{\small
\begin{align}
     &\langle \nabla \left ( \mathcal L(\mathbf{w}_{1})- \frac{\varsigma}{2}\left \| \mathbf{w}_{1} \right \|_{2}^{2}\right )-\nabla \left ( \mathcal L(\mathbf{w}_{2})- \frac{\varsigma}{2}\left \| \mathbf{w}_{2} \right \|_{2}^{2}\right ) , \mathbf{w}_{1}  - \mathbf{w}_{2} \rangle\notag\\ 
     &  \geq \frac{1}{L+\mu}\left \| \nabla \left ( \mathcal L(\mathbf{w_{1}})- \frac{\varsigma}{2}\left \| \mathbf{w}_{1} \right \|_{2}^{2}\right ) -\nabla \left ( \mathcal L(\mathbf{w}_{2})- \frac{\varsigma}{2}\left \| \mathbf{w}_{2} \right \|_{2}^{2}\right )\right \|_{2}^{2}
\end{align}
}

\noindent And therefore
{\small
\begin{align}
     & \langle \nabla\mathcal L(\mathbf{w}_{1})- \nabla\mathcal L(\mathbf{w}_{2}), \mathbf{w}_{1} - \mathbf{w}_{2} \rangle -  \langle \varsigma \mathbf{w}_{1}- \varsigma\mathbf{w}_{2}, \mathbf{w}_{1} - \mathbf{w}_{2} \rangle \notag\\ 
     &  \geq \frac{1}{L-\varsigma}\left \| \left ( \nabla\mathcal L(\mathbf{w}_{1})- \nabla\mathcal L(\mathbf{w}_{2})  \right ) - \left ( \varsigma \mathbf{w}_{1}- \varsigma\mathbf{w}_{2} \right )\right \|_{2}^{2}
\end{align}
}

\noindent Refer to Assumption \ref{as:assumption1}, we obtain
{\small
\begin{align}
    &\langle \nabla\mathcal L(\mathbf{w}_{1})- \nabla\mathcal L(\mathbf{w}_{2}), \mathbf{w}_{1}  - \mathbf{w}_{2} \rangle  \geq  \frac{L\varsigma}{L-\varsigma}\left \|\mathbf{w}_{1}-\mathbf{w}_{2}\right \|_{2}^{2}\notag\\
   & - \frac{2\varsigma}{L-\varsigma}\left \langle \nabla \mathcal L(\mathbf{w}_{1}) - \nabla \mathcal L(\mathbf{w}_{2}), \mathbf{w}_{1}-\mathbf{w}_{2}  \right \rangle \notag\\
    & +\frac{1}{L-\varsigma}\left \| \nabla \mathcal L(\mathbf{w}_{1}) - \nabla \mathcal L(\mathbf{w}_{2})\right \|_{2}^{2} \notag\\
    & \geq  -\frac{L\varsigma}{L-\varsigma}\left \|\mathbf{w}_{1}-\mathbf{w}_{2}\right \|_{2}^{2} +\frac{1}{L-\varsigma}\left \| \nabla \mathcal L(\mathbf{w}_{1}) - \nabla \mathcal L(\mathbf{w}_{2})\right \|_{2}^{2} 
\end{align}
}

\noindent Let $\varsigma = -\mu$, then we conclude the proof of Lemma \ref{le:lemma1}.

\subsubsection{Proof of Lemma \ref{le:lemma2}}
%\proofname
We have the following equation:
{\small
\begin{align}
     \left \|\mathbf{m}(\mathbf{w})-\nabla \mathcal L(\mathbf{w})  \right \|_{2} & \leq  \left \| \mathbf{m}(\mathbf{w})-\mathbf{m}_{0}(\mathbf{w})\right \|_{2} \notag\\
     & +\left \| \mathbf{m}_{0}(\mathbf{w})- \nabla \mathcal L(\mathbf{w})\right \|_{2} 
\end{align}
}

\noindent from~\cite{fu2019attack} inequality 18, we know $\forall i, n, \exists E > 0$  
$$ \sup \left | e_{i,n} \right| \leq \frac{E}{\sqrt{M}\delta}$$
Where 
{\small $$E = \sup \left \{ \frac{37\sqrt{2}\lambda(M+4)}{25(M-1)}
\underset{i}{\mathrm{median}} \left\{ |w_{i,n}^{t}-\hat{B}_{n}x_{i,n}^{t}-\hat{A}_{n}|\right \} \right \}$$}

\noindent and the dimension of $\mathbf{w}$ is $N$. Hence the distance between the two aggregation functions satisfies 
{\small
\begin{align}
    \left \| \mathbf{m}(\mathbf{w})-\mathbf{m}_{0}(\mathbf{w})\right \|_{2} \leq \sqrt{N}\left \|  \sum_{i=1}^{M}\bar{R}_{i}\left ( \hat{B}_{i}\left ( M-1 \right )+\frac{2E}{\sqrt{M}\delta}  \right ) \right \|_{2}
\end{align}
}

\noindent Based on Equation~\ref{eq:final reputation}
{\small
\begin{align}
\frac{s\exp(-cs)}{\sum_{j=0}^{s}\exp(-cj)}&\cdot\frac{Wa}{\eta N+W} 
  \leq \tilde{R_{i}^{t}} 
\end{align}
}
{\small
\begin{align}
    \tilde{R_{i}^{t}} 
 \leq \frac{s}{\sum_{j=0}^{s}\exp(-cj)}
\cdot\frac{\kappa N + Wa}{\kappa N+W} 
\end{align}
}

\noindent so we have
{\small
\begin{align}
    \bar{R}_{i} = \frac{\tilde{R_{i}^{t}}}{\sum_{i=1}^{M}\tilde{R_{i}^{t}}} \leq \frac{1}{\frac{Wa(M-1)(\kappa N + W)}{(\eta N + W)(\kappa N + Wa)}+1}
\end{align}
}

\noindent Due to our Aggregation Algorithm%~\ref{al:algo1},
{\small
\begin{align}
   \hat{B}_{n} = \underset{i}{\mathrm{median}}\left ( {\underset{i\neq j} {\mathrm{median}}\frac{w_{j,n}-w_{i,n}}{x_{j,n}-x_{i,n}}} \right ) \leq \varpi
\end{align}
}

\noindent Therefore, we have
{\small
\begin{align}
    \left \|  \sum_{i=1}^{M}\bar{R}_{i}\left ( \hat{B}_{i}\left ( M-1 \right )+\frac{2E}{\sqrt{M}\delta} \right ) \right \|_{2} \leq & \frac{M(\varpi(M-1)+\frac{2E}{\sqrt{M}\delta})}{\frac{Wa(M-1)(\kappa N + W)}{(\eta N + W)(\kappa N + Wa)}+1} \notag\\
    & = \Delta_{1}
\end{align}
}

\noindent Hence, we proof Lemma~\ref{le:lemma2}.

\subsubsection{Proof of Theorem~\ref{th:theorem1}}
We first consider a general problem of robust estimation of a one dimensional random variable. Suppose that there are $M$ clients, and $p$ percentage of them are malicious and own adversarial data. %which is $q$ percentage of the whole training data.
In $t$ iteration, we have:
{\small
\begin{align}
\label{eq:loss function}
     & \left \|\mathbf{w}^{t} - \mathbf{w}^{*} \right \|_{2}  = \left \| (\mathbf{w}^{t-1}-r\mathbf{m}(\mathbf{w}^{t-1})- \mathbf{w}^{*}\right \|_{2} \notag\\ 
     & \leq \underset{A}{\underbrace{\left \| \mathbf{w}^{t-1} - r\nabla\mathcal{L}(\mathbf{w}^{t-1}) - \mathbf{w}^{*} \right \|_{2}}} \notag\\
     & + \underset{B}{\underbrace{r\left \|\mathbf{m}(\mathbf{w}^{t-1}) - \nabla\mathcal{L}(\mathbf{w}^{t-1})\right \|_{2}}}
\end{align}
}

\noindent We bound part $A$ first. We have 
{\small
\begin{align}
\label{eq:equation20}
    &\left \| \mathbf{w}^{t-1} -  r\nabla\mathcal{L}(\mathbf{w}^{t-1}) - \mathbf{w}^{*} \right \|_{2}^{2}  =  \left \| \mathbf{w}^{t-1} - \mathbf{w}^{*} \right \|_{2}^{2} \notag\\
    &  + r^{2} \left \| \nabla\mathcal{L}(\mathbf{w}^{t-1}) \right \|_{2}^{2}
    - 2r\left \langle \nabla \mathcal L(\mathbf{w}^{t-1}),\mathbf{w}^{t-1} - \mathbf{w}^{*} \right \rangle 
\end{align}
}

\noindent Under the Assumption \ref{as:assumption4} and Lemma \ref{le:lemma1}, we have
{\small
\begin{align}
\label{eq:equation21}
\left \langle  \nabla{\mathcal L(\mathbf{w}^{t-1}), \mathbf{w}^{t-1}- \mathbf{w^{*}}}\right \rangle 
 &\geq  \frac{L\mu}{L+\mu}\left \| \mathbf{w}^{t-1}-\mathbf{w^{*}}\right \|_{2}^{2} \notag \\
 & +\frac{1}{L+\mu}\left \| \nabla \mathcal L( \mathbf{w}^{t-1})\right \|_{2}^{2}
\end{align}
}

\noindent Then we combine inequalities~\ref{eq:equation21} to equation~\ref{eq:equation20}
{\small
\begin{align}
\label{eq:equation22}
     \left \| \mathbf{w}^{t-1} - r\nabla\mathcal{L}(\mathbf{w}^{t-1}) - \mathbf{w}^{*} \right \|_{2}^{2}  &\leq (1-2r\frac{L\mu}{L+\mu})\left \| \mathbf{w}^{t-1} - \mathbf{w}^{*} \right \|_{2}^{2}\notag\\
     & + (r^{2}-\frac{2r}{L+\mu})\left \| \nabla \mathcal L( \mathbf{w}^{t-1})\right \|_{2}^{2}
\end{align}
}

\noindent From Assumption~\ref{as:assumption1}, we can derive:
{\small
\begin{align}
\label{eq:equation23}
    \left \| \nabla \mathcal L( \mathbf{w}^{t-1}) - \nabla \mathcal L( \mathbf{w}^{*})\right \|_{2}^{2} \leq L^2 \left \| \mathbf{w}^{t-1} - \mathbf{w}^{*} \right \|_{2}^{2}
\end{align}
}

\noindent Combining inequalities~\ref{eq:equation22} and~\ref{eq:equation23}, we have:
{\small
\begin{align}
\label{eq:equation24}
\left \| \mathbf{w}^{t-1} - r\nabla\mathcal{L}(\mathbf{w}^{t-1}) - \mathbf{w}^{*} \right \|_{2}^{2}  \leq (1-Lr)^2\left \| \mathbf{w}^{t-1} - \mathbf{w}^{*} \right \|_{2}^{2}
\end{align}
}

\noindent Let $r < \frac{1}{L}$, we have
{\small
\begin{align}
\label{eq:equation25}
\left \| \mathbf{w}^{t-1} - r\nabla\mathcal{L}(\mathbf{w}^{t-1}) - \mathbf{w}^{*} \right \|_{2}  \leq \left ( 1- Lr \right )\left \| \mathbf{w}^{t-1} - \mathbf{w}^{*} \right \|_{2}
\end{align}
}

Then we turn to bound part $B$. Based on Lemma~\ref{le:lemma2}, we know:
{\small
\begin{align}
     \left \|\mathbf{m}(\mathbf{w})-\nabla \mathcal L(\mathbf{w})  \right \|_{2} & \leq  \left \| \mathbf{m}_{0}(\mathbf{w})- \nabla \mathcal L(\mathbf{w})\right \|_{2} + \sqrt{N}\Delta_{1}
\end{align}
}

Assume Assumption~\ref{as:assumption1}, \ref{as:assumption2}, \ref{as:assumption3} and~\ref{as:assumption4} holds, and $\exists \epsilon $ fulfills inequality~\ref{ineq:probablity}. Based on Lemma 1 in~\cite{yin2018byzantine}, with the probability $1- \xi \geq 1- \frac{4d}{\left ( 1+\hat{Q}ML\upsilon \right)^{d}}$, we have
{\small
\begin{align}
    \left \| \mathbf{m}_{0}(\mathbf{w})- \nabla \mathcal L(\mathbf{w})\right \|_{2} \leq & \sqrt{\frac{2}{\hat{Q}}} D_{\epsilon}V_{w} (\sqrt{\frac{d\log(1+\hat{Q}ML\upsilon)}{M(1-p)}}\notag\\
    & +C\frac{\mathcal{G}_{w}}{\sqrt{\hat{Q}}}+p) +2\sqrt{2}\frac{1}{M\hat{Q}}=  \Delta_{2}
\end{align}
}

\noindent where $C=0.4748$. After obtaining the bound of part $A$ and $B$, now we have
{\small
\begin{align}
\label{eq:loss_boundary}
     & \left \|\mathbf{w}^{t} - \mathbf{w}^{*} \right \|_{2} \leq \underset{Bound\;A}{\underbrace{ \left(1- Lr \right) \left \| \mathbf{w}^{t-1} - \mathbf{w}^{*} \right \|_{2}}} + \underset{Bound\;B}{\underbrace{r\left(\sqrt{N}\Delta_{1} + \Delta_{2}\right)}}
\end{align}
}

Hence, we can prove Theorem~\ref{th:theorem1} through iterations using the formula of a finite geometric series,
{\small
\begin{align}
     \left \| \mathbf{w}^{t} - \mathbf{w}^{*} \right \|_{2} \leq & \left ( 1- Lr \right )^{t}\left \| \mathbf{w}^{0} - \mathbf{w}^{*} \right \|_{2}\notag \\
    & + \frac{1-(1-Lr)^{t}}{Lr}r\left(\sqrt{N}\Delta_{1} + \Delta_{2}\right)\notag\\
    & \leq  \left ( 1- Lr \right )^{t}\left \| \mathbf{w}^{0} - \mathbf{w}^{*} \right \|_{2}
     + \frac{1}{L}\left(\sqrt{N}\Delta_{1} + \Delta_{2}\right)
\end{align}
}

\subsection{Experimental Setting}\label{app:experimental_setting}
Our simulation experiments are implemented with Pytorch framework~\cite{paszke2017automatic} on the cloud computing platform Google Colaboratory Pro (Colab Pro) with access to Nvidia K80s, T4s, P4s and P100s with 25 GB of Random Access Memory. Table~\ref{tab:setting} shows the default setting in our experiments.
%For the training setting, the local models are trained independently and sequentially in each round of training before being aggregated into the global model. Also we presume that clients do not leave the system. 
\begin{table}[!hbpt]
    \caption{Default experimental settings}
    % \vspace{-3mm}
{%
\centering
     \begin{tabular}{lll}
        \toprule
        Explanation & Notation & Default Setting\\
        \midrule
        prior probability & $a$ & 0.5 \\ 
        non-information prior weight & $W$ & 2 \\ 
        weight for positive observation & $\kappa$ & 0.3 \\ 
        time decay parameter & $c$ & 0.5 \\
        window length & $s$ & 10 \\
        confidence threshold & $\delta$ & 0.1 \\
        value range & $\varpi$ & 2 \\
        Objective Function & $\mathcal{L}(\mathbf{\cdot})$ &  Negative Log-likelihood Loss\\  
        \midrule
         Learning rate & $r$ & 0.01\\
         Batch size per client & & 64\\
        The number of local iterations & & 10 \\
        The number of total iterations & & 100 \\
      \bottomrule
    \end{tabular}
    }\\~\\
    \label{tab:setting}
\end{table}

\subsection{Supplementary dataset for experiment}\label{app:CIFAR-10}
CIFAR-10 is a supplementary dataset  assessing the robustness of our reputation scheme in image datasets to poisoning attacks. The CIFAR-10 dataset is a $32\times32$ colour image dataset that includes ten classes with a total number of 50 thousand images for training and 10 thousand images for testing. Here we use ResNet-18~\cite{he2016deep} model pertaining to ImageNet~\cite{deng2009imagenet} with 20 iterations. In CIFAR-10 dataset, attackers switch the label of ``cat'' images to the ``dog''.

\begin{figure*}[!hbpt]
     \centering
      \includegraphics[width=1\textwidth]{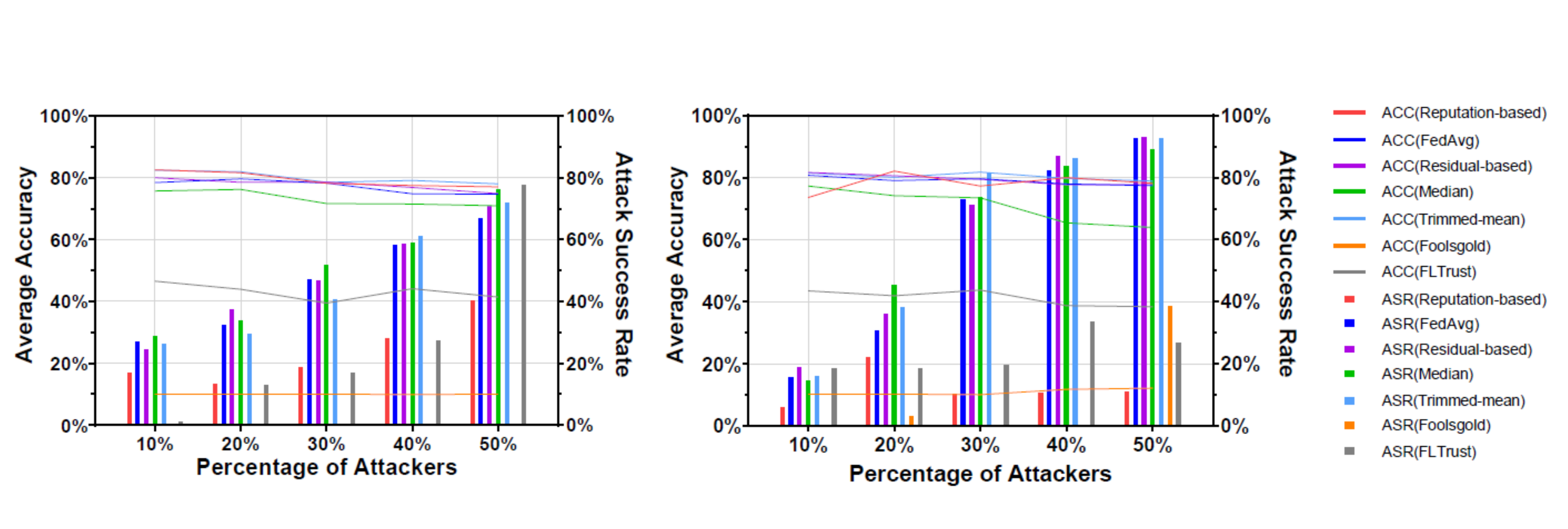}
     \caption{Average accuracy (ACC) and attack success rate (ASR) for varying percentage of attackers from 10\% to 50\% under label flipping (left) and backdoor (right) attack for Reputation, FedAvg, Residual-based, Median, Trimmed-Median, FoolsGold and FLTruts in CIFAR-10 dataset.
     %and (b) CIFAR-10 Dataset
     }
     \label{fig:CIFAR_attack}
\end{figure*}

Figure~\ref{fig:CIFAR_attack} shows that under label flipping and backdoor attack during the whole process, our reputation-based method has the highest accuracy outperforming other methods and the lowest ASR, excluding Foolsgold in CIFAR-10, yielding a result that is similar to the one in SURL.

\begin{figure*}[!hbpt]
     \centering
      \includegraphics[width=1\textwidth]{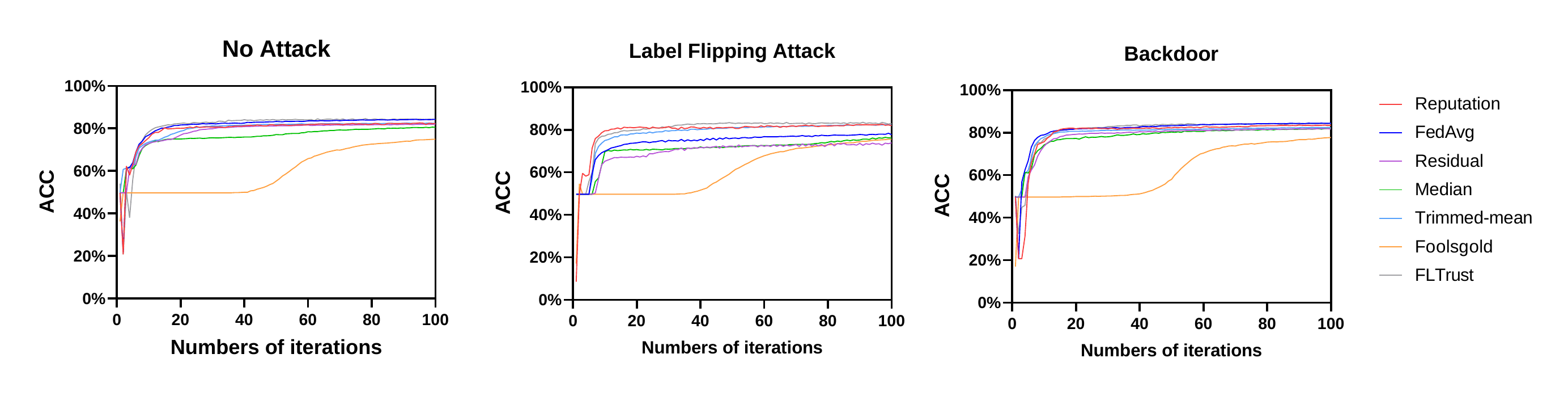}
     \caption{Average accuracy (ACC) with no attack (left) and varying percentage of attackers from 10\% to 50\% under label flipping (middle) and backdoor (right) attack for Reputation, FedAvg, Residual-based, Median, Trimmed-mean, Foolsgold, FLTrust in SURL dataset.}
     \label{fig:convergence_100}
\end{figure*}

\begin{figure*}[!hbpt]
     \centering
      \includegraphics[width=1\textwidth]{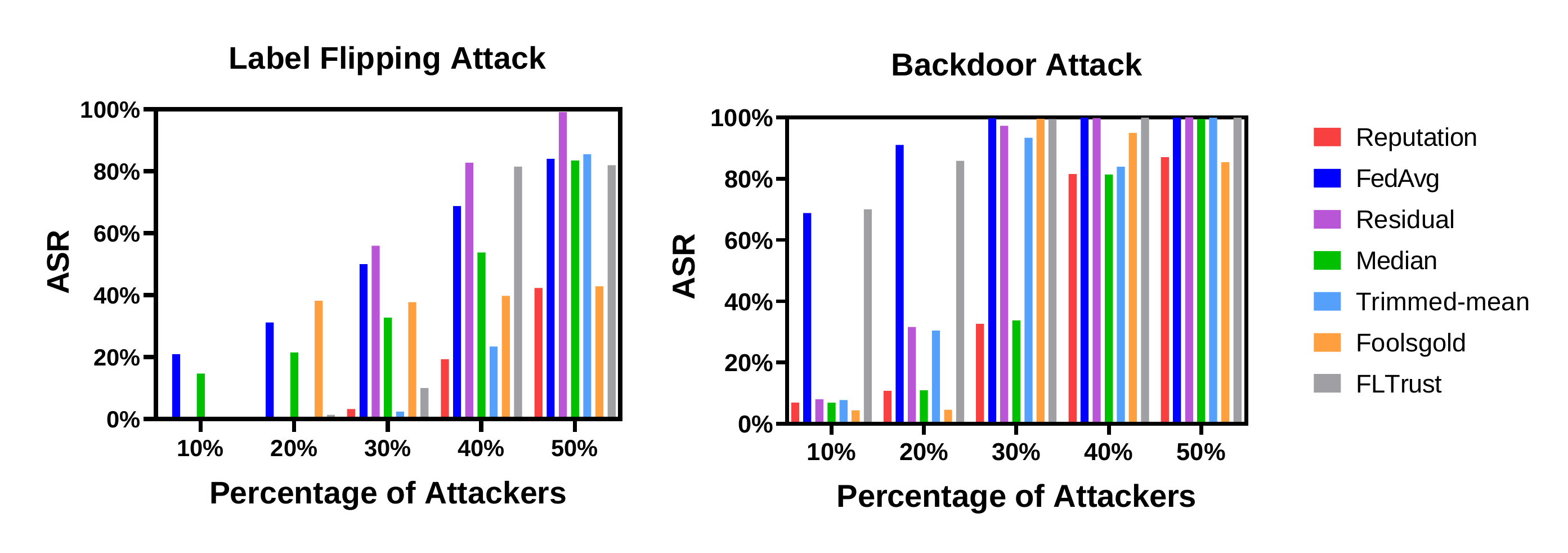}
     \caption{Attack success rate (ASR) for varying percentage of attackers from 10\% to 50\% under label flipping (left) and backdoor (right) attack for Reputation, FedAvg, Residual-based, Median, Trimmed-mean, Foolsgold, FLTrust in SURL dataset.
     }
     \label{fig:attack_100}
\end{figure*}

\subsection{Extra Experimental Result}

We evaluate the proposed defence for a varying number of clients from 10 to 200 in the SURL dataset. Here, we analyze the performance of the aforementioned methods for 100 clients. The results are presented in Figure~\ref{fig:convergence_100} and Figure~\ref{fig:attack_100} that correspond to the average accuracy (ACC) and attach success rate (ASR), respectively.

In the no attack scenario, we observe that our method converges between 1.7$\times$ to 7.1$\times$ faster than all competing state-of-the-art methods with at least as good performance (or outperforms) compared with competing methods in terms of classification accuracy, see Figure~\ref{fig:convergence_100} (left). In addition, even under the two different attacks, our method:
     (i) converges  between 1.6$\times$ to 3.6$\times$ faster than all competing state-of-the-art methods,
    (ii) provides the same or better accuracy than competing methods, 
    and (iii) yields the lowest ASR compared to all other methods.

\end{document}